\documentclass{pasj00}
\color{black}
\tenpoint

\SetRunningHead{K. Sato et al.}{
Suzaku observation of Abell~262}

\title{
Temperature and Metallicity \\
in the Intra-cluster Medium of ABELL~262 \\
observed with Suzaku}
\author{
 Kosuke \textsc{Sato},\altaffilmark{1,2}
 Kyoko \textsc{Matsushita},\altaffilmark{1}
and Fabio \textsc{Gastaldello},\altaffilmark{3,4,5} 
}
\altaffiltext{1}{
Department of Physics, Tokyo University of Science,
1-3 Kagurazaka, Shinjuku-ku, Tokyo 162-8601}
\altaffiltext{2}{
Graduate School of Natural Science and Technology,
Kanazawa University, 
Kakuma, Kanazawa, Ishikawa 920-1192}
\email{ksato@astro.s.kanazawa-u.ac.jp}
\altaffiltext{3}{IASF-Milano, INAF, via Bassini 15, Milano 20133, Italy}
\altaffiltext{4}{Occhialini Fellow}
\altaffiltext{5}{Department of Physics and Astronomy, University of
California at Irvine, 4129 Frederick Reines Hall, Irvine, CA 92697-4575}
\Received{2008~August~1}
\Accepted{2008~October~21}
\Published{---}

\begin{document}
\maketitle

\begin{abstract}
We studied the temperature and abundance distributions of
intra-cluster medium (ICM) in the Abell~262 cluster of galaxies
observed with Suzaku.  Abell ~262 is a bright, nearby poor cluster
with ICM temperature of $\sim2$ keV, thus providing useful information
about the connection of ICM properties between groups and clusters of
galaxies.  The observed spectrum of the central region was
well-represented by two temperature models, and the spectra for the
outer regions were described by single temperature model. With the XIS
instrument, we directly measured not only Si, S, and Fe lines but also
O and Mg lines and obtained those abundances to an outer region of
$\sim0.43~ r_{180}$ for the first time.  We found steep gradients for
Mg, Si, S, and Fe abundances, while O showed almost flat abundance
distribution.  Abundance ratios of $\alpha$-elements to Fe were found
to be similar to those of the other clusters and groups.  We
calculated the number ratio of type II to type Ia supernovae for the
ICM enrichment to be $3.0\pm 0.6$ within $0.1 r_{180}$, and the value
was consistent with those for other clusters and groups.  We also
calculated metal mass-to-light ratios (MLRs) for Fe, O and Mg (IMLR,
OMLR, MMLR) with B-band and K-band luminosities of the member galaxies
of Abell~262. The derived MLRs were comparable to those for other
clusters with $kT =$ 3--4 keV\@.
  
\end{abstract}

\section{Introduction}

The metal abundances of Intra-cluster medium (ICM) carry
a lot of information in understanding
the chemical history and evolution of clusters.
X-ray observations of ICM can immediately
determine its temperature and metal abundances. 
Recent X-ray observations allow us to measure temperature and metal 
abundance distributions in the ICM based on the spatially resolved spectra.  
A large amount of metals of the ICM are mainly produced by supernovae (SNe) 
in early-type galaxies \citep{arnaud92,renzini93}, 
which are classified roughly as type Ia (SNe~Ia) and type II (SNe~II)\@.
In order to know how the ICM has been
enriched, we need to measure the amount and distribution of metals in
the ICM\@.  Because Si and Fe are both synthesized in SNe Ia and II, 
we need to know O and Mg abundances,
which are synthesized predominantly in SNe II, in resolving the past
metal enrichment process in ICM by supernovae.  

ASCA firstly showed the distribution of the heavy elements, 
such as Si and Fe, in the ICM
\citep{fukazawa98,fukazawa00,finoguenov00,finoguenov01}, and
BeppoSax also showed the Fe distribution and its gradient 
\citep{degrandi01}.
\citet{renzini97} and \citet{makishima01} summarized
iron-mass-to-light ratios (IMLR) with B-band luminosity 
for various objects, as a function of
their plasma temperature serving as a measure of the system richness, 
and IMLRs in groups were found to be smaller than those in clusters. 
They also showed that the early-type galaxies released large
amount of metals which were probably formed through past supernovae 
explosions as shown earlier by \citet{arnaud92}.  
In order to obtain a correct modeling of ICM, 
we need to know the correct temperature and metal abundance profiles
without biases (e.g., \cite{buote00,sanders02}).
Especially for the ICM of cooler systems, such as elliptical galaxies and 
groups of galaxies, sensitive observations are required as mentioned 
in \citet{arimoto97} and \citet{matsushita00}.

Recent XMM-Newton and Chandra observations have enabled us to study
the temperature structure, and detailed properties of the heavy
elements in the ICM were shown based on the high spatial resolutions
and small point spread functions
\citep{fabian01,fabian05,fabian06,finoguenov02,xu02,gastaldello02,
matsushita03,matsushita07b,tamura03,sanders07,humphrey06,rasmussen07}.
In addition, the spatial distribution and elemental abundance pattern
of the ICM metals were determined with the large effective area of
XMM-Newton
\citep{matsushita07b,tamura04,boehringer05,osullivan05,sanders06,
deplaa06,deplaa07,werner06,simionescu08}.  However, the abundance
measurements of O and Mg with XMM-Newton, in particular for the outer
regions of clusters, are quite difficult due to the relatively high
intrinsic background. Suzaku XIS can measure all the main elements
from O to Fe, because it realizes lower background level and higher
spectral sensitivity, especially below 1 keV \citep{koyama07}.  Suzaku
observations have shown the abundance profiles of O, Mg, Si, S, and Fe
to the outer regions with good precision for several clusters
\citep{matsushita07a,sato07a,sato08a,sato08b,tokoi08}.  Comparing the
Suzaku results with supernova nucleosynthesis models, \citet{sato07b}
showed the number ratios of SNe~II to Ia to be ~3.5\@.

Abell~262 (hereafter A~262) is a nearby cluster of galaxies
($z=0.0163$) characterized by a smooth and symmetric distribution of
ICM, and is a richness class 0 cluster.  A~262 is located on a knot of
the large-scale filament of the Pisces-Perseus supercluster.  A~262 is
a very useful system in studying the history of chemical evolution of
groups and clusters, because the ICM temperature is about $\sim2$ keV
which is intermediate between clusters and groups of galaxies.  The
central cD galaxy, NGC~708, is a host of a double-lobed radio source,
B20149+35 \citep{parma86,fanti86}.  In the past, Einstein
\citep{david93}, ROSAT \citep{david96,peres98,neill01}, ASCA
\citep{white00,fukazawa04}, XMM-Newton \citep{peterson03}, and Chandra
\citep{blanton04} observed A~262, and ASCA showed an average
temperature of $kT=2.13$ keV and an average metal abundance of 0.44
solar, respectively \citep{fukazawa04}.
In a future paper \citep{gastaldello08} we will 
investigate the mass and entropy radial profiles.

This paper reports on results from Suzaku observations of A~262
out to $27'\simeq 540\; h_{70}^{-1}$~kpc, corresponding to
$\sim 0.43\; r_{180}$.  We use $H_0=70$
km~s$^{-1}$~Mpc$^{-1}$, $\Omega_{\Lambda} = 1-\Omega_M = 0.73$ in this
paper.  At a redshift of $z=0.0163$, $1'$ corresponds to 19.9~kpc,
and the virial radius, $r_{\rm 180} = 1.95\;
h_{100}^{-1}\sqrt{k\langle T\rangle/10~{\rm keV}}$~Mpc
\citep{markevitch98}, is 1.25~Mpc ($63'$) for an average temperature
of $k\langle T\rangle = 2.0$~keV\@.  Throughout this paper we adopt
the Galactic hydrogen column density of $N_{\rm H} = 5.38\times
10^{20}$ cm$^{-2}$ \citep{dickey90} in the direction of A~262\@.
Unless noted otherwise, the solar abundance table is given by
\citet{anders89}, and the errors are in the 90\% confidence region 
for a single interesting parameter.

\begin{table*}
\caption{Suzaku Observation logs for Abell~262.}
\label{tab:1}
\begin{tabular}{lccccc} \hline 
Region & Seq. No. & Obs. date & \multicolumn{1}{c}{(RA, Dec)$^\ast$} &Exp.&After screening \\
&&&J2000& ksec &(BI/FI) ksec \\
\hline 
center & 802001010 & 2007-08-17T04:13:43 & (\timeform{01h52m46.1s},
 \timeform{+36D09'33''})& 37.2& 37.0/37.0\\
offset1 & 802079010 & 2007-08-06T21:27:38 & (\timeform{01h54m08.3s},
 \timeform{+36D16'07''})& 54.6& 54.3/54.3\\
offset2 & 802080010 & 2007-08-08T01:52:10 & (\timeform{01h52m14.4s},
 \timeform{+36D26'08''})& 54.7& 54.3/54.4\\
\hline\\[-1ex]
\multicolumn{6}{l}{\parbox{0.9\textwidth}{\footnotesize 
\footnotemark[$\ast$]
Average pointing direction of the XIS, written in the 
RA\_NOM and DEC\_NOM keywords of the event FITS files.}}\\
\end{tabular}
\end{table*}

\begin{figure}
\centerline{
\FigureFile(0.5\textwidth,0.5\textwidth){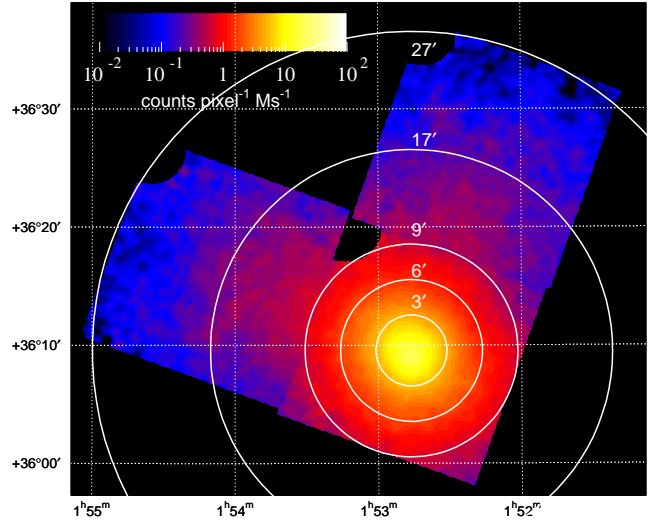}}
\caption{
Combined XIS image of central observation in the
0.5--4.0 keV energy range. The observed XIS0, 1, and 3 images 
were added on the sky coordinate after removing each calibration source region,
and smoothed with $\sigma=16$ pixel $\simeq 17''$ Gaussian.
Estimated components of extragalactic X-ray background (CXB)
and instrumental background (NXB) were subtracted,
and the exposure was corrected, though vignetting was not corrected.
The white circles show the extracted regions in the spectral fits.
}\label{fig:1}
\end{figure}

\section{Observations and Data Reduction}\label{sec:obs}

\subsection{Observation}
\label{subsec:obs}

Suzaku observed the central and offset regions of Abell~262 in August 2007
(PI:~F. Gastaldello and K. Matsushita).  The observation log
is given in table~\ref{tab:1}, and the XIS image in 0.5--4~keV is
shown in figure~\ref{fig:1}.  We analyzed only the XIS data in this
paper, although Suzaku observed the object with both XIS and HXD\@.
The XIS instrument consists of three sets 
of X-ray CCDs (XIS~0, 1, and 3). XIS~1 is a back-illuminated (BI) 
sensor, while XIS~0 and 3 are front-illuminated (FI)\@.  The 
instrument was operated in the Normal clocking mode (8~s exposure 
per frame), with the standard $5\times 5$ or $3\times 3$ editing mode.

\subsection{Data Reduction}

We used version 2.0 processing data, and the analysis was performed 
with HEAsoft version 6.4.1 and XSPEC 11.3.2aj.
Details of the analysis method are given in \citet{sato07a}.
Here we give just a brief description of the data reduction.
The light curve of each sensor in the 0.3--10~keV range with a 16~s time
bin was also examined in order to exclude periods with anomalous event
rates which were greater or less than $\pm 3\sigma$ around the mean to 
remove the charge exchange contamination \citet{fujimoto07}, 
while Suzaku data was little affected by the soft proton flare compared 
with XMM data.
The exposure after the screening was essentially the same as that
before screening in table~\ref{tab:1}, which indicated that the non
X-ray background (NXB) was almost stable during the observation. 
Event screening with cut-off rigidity (COR) was not performed in our data.
In order to subtract the NXB and the extra-galactic
cosmic X-ray background (CXB), we employed the dark Earth
database by the ``xisnxbgen'' Ftools task, 
and employed the CXB spectrum given by \citet{kushino02}.

We generated two Ancillary Response Files (ARFs) for the spectrum 
of each annular sky region, $A^{\makebox{\small\sc u}}$ and 
$A^{\makebox{\small\sc b}}$, which respectively assumed uniform sky 
emission and $\sim 1^{\circ} \times 1^{\circ}$ size of the $\beta$-model 
surface brightness profile, $\beta = 0.38$ and $r_c =$ \timeform{0.47'}, 
in \citet{fukazawa04}, by the ``xissimarfgen'' ftools task \citep{ishisaki07}.
We also included the effect of the contaminations on the optical blocking 
filter of the XISs in the ARFs.
Since the energy resolution also slowly degraded
after the launch, due to radiation damage, this effect was included
in the Redistribution Matrix File (RMF) by the ``xisrmfgen'' Ftools task.

\begin{table*}
\caption{
Area, coverage of whole annulus, {\scriptsize SOURCE\_RATIO\_REG},
and observed/estimated counts for each annular region.
}\label{tab:2}
\centerline{
\begin{tabular}{lrrrcrrrrcrrrr}
\hline\hline
\makebox[3em][l]{Region\,$^\ast$} & \multicolumn{1}{c}{Area\makebox[0in][l]{\,$^\dagger$}} & Coverage\makebox[0in][l]{\,$^\dagger$}\hspace*{-0.5em} & \makebox[4.2em][r]{\scriptsize SOURCE\_\makebox[0in][l]{\,$^\ddagger$}}\hspace*{-0.5em} & Energy & \multicolumn{4}{c}{BI counts\makebox[0in][l]{\,$^\S$}} &$\!\!\!\!$& \multicolumn{4}{c}{FI counts\makebox[0in][l]{\,$^\S$}} \\
\cline{6-9}\cline{11-14}
& \makebox[2em][c]{(arcmin$^2$)} &      & \makebox[4.2em][r]{\scriptsize RATIO\_REG}\hspace*{-0.5em} & (keV) & OBS & NXB & CXB & $f_{\rm BGD}$ &$\!\!\!\!$& OBS & NXB & CXB & $f_{\rm BGD}$ \\
\hline\\[-2ex]
Center&&&&&&&&&&&&&\\
0$'$--3$'$& 28.7&100.0\%& 13.6\%& 0.4--7.1&$\!\!39,224$&$\!\!  316$&$\!\!338$&$\!\! 1.7\%$&$\!\!\!\!$&$\!\!57,652$&$\!\!336$&$\!\!534$&$\!\! 1.5\%$\\
3$'$--6$'$&84.8&100.0\%& 11.7\%& 0.4--7.1&$\!\!37,674$&$\!\!953$&$\!\!871$&$\!\!4.8\%$&$\!\!\!\!$&$\!\!52,596$&$\!\!970$&$\!\!1,371$&$\!\!4.5\%$\\
6$'$--9$'$&133.4&94.3\%& 9.8\%& 0.4--7.1&$\!\!24,725$&$\!\!1,551$&$\!\!1,138$&$\!\!10.9\%$&$\!\!\!\!$&$\!\!33,040$&$\!\!1,510$&$\!\!1,705$&$\!\!9.7\%$\\
\hline
Offset1&&&&&&&&&&&&&\\
9$'$--17$'$ & 121.2& 18.5\%&  3.9\%& 0.4--7.1&$\!\! 13,956$&$\!\!1,704$&$\!\!1,448$&$\!\!22.6\%$&$\!\!\!\!$&$\!\! 18,576$&$\!\!1,922$&$\!\!2,463$&$\!\!23.6\%$\\
17$'$--27$'$ &161.8& 11.7\%&  3.3\%& 0.4--7.1&$\!\!11,585$&$\!\!2,718$&$\!\!2,314$&$\!\!43.4\%$&$\!\!\!\!$&$\!\!13,507$&$\!\!2,514$&$\!\!3,398$&$\!\!43.8\%$\\
\hline
Offset2&&&&&&&&&&&&&\\
9$'$--17$'$ & 119.9& 18.3\%&  4.6\%& 0.4--7.1&$\!\! 12,625$&$\!\!1,945$&$\!\!1,582$&$\!\!27.9\%$&$\!\!\!\!$&$\!\! 15,062$&$\!\!1,896$&$\!\!2,268$&$\!\!27.6\%$\\
17$'$--27$'$ &162.6& 11.8\%&  3.1\%&0.4--7.1&$\!\!9,884$&$\!\!2,667$&$\!\!2,244$&$\!\!49.8\%$&$\!\!\!\!$&$\!\!12,367$&$\!\!2,640$&$\!\!3,571$&$\!\!50.2\%$\\
\hline
\end{tabular}
}

\medskip
\parbox{\textwidth}{\footnotesize
{\scriptsize SOURCE\_RATIO\_REG} represents the
flux ratio in the assumed spatial distribution on the sky
($\beta$-model) inside the accumulation region
to the entire model, and written in the header keyword of
the calculated ARF response by ``xissimarfgen''.
}
\parbox{\textwidth}{\footnotesize
\footnotemark[$\ast$]
The inner three regions correspond to the central observation, 
and the outer two regions correspond to the offset observations.

\footnotemark[$\dagger$]
The average values among four sensors are presented.

\footnotemark[$\ddagger$]
$\makebox{\scriptsize\rm SOURCE\_RATIO\_REG}\equiv
\makebox{\scriptsize\rm COVERAGE}\;\times
\int_{r_{\rm in}}^{r_{\rm out}} S(r)\; r\,dr / 
\int_{0}^{\infty} S(r)\; r\,dr$,
where $S(r)$ represents the assumed radial profile
of Abell~262, and we defined $S(r)$ in $1^{\circ}\times 1^{\circ}$ 
region on the sky.

\footnotemark[$\S$]
OBS denotes the observed counts including NXB and CXB in
 0.4--7.1~keV\@. 
NXB and CXB are the estimated counts.
}
\end{table*}

\section{Temperature and Abundance Profiles}

\subsection{Spectral Fit}
\label{sec:spec}

\begin{figure*}
\begin{minipage}{0.33\textwidth}
\FigureFile(\textwidth,\textwidth){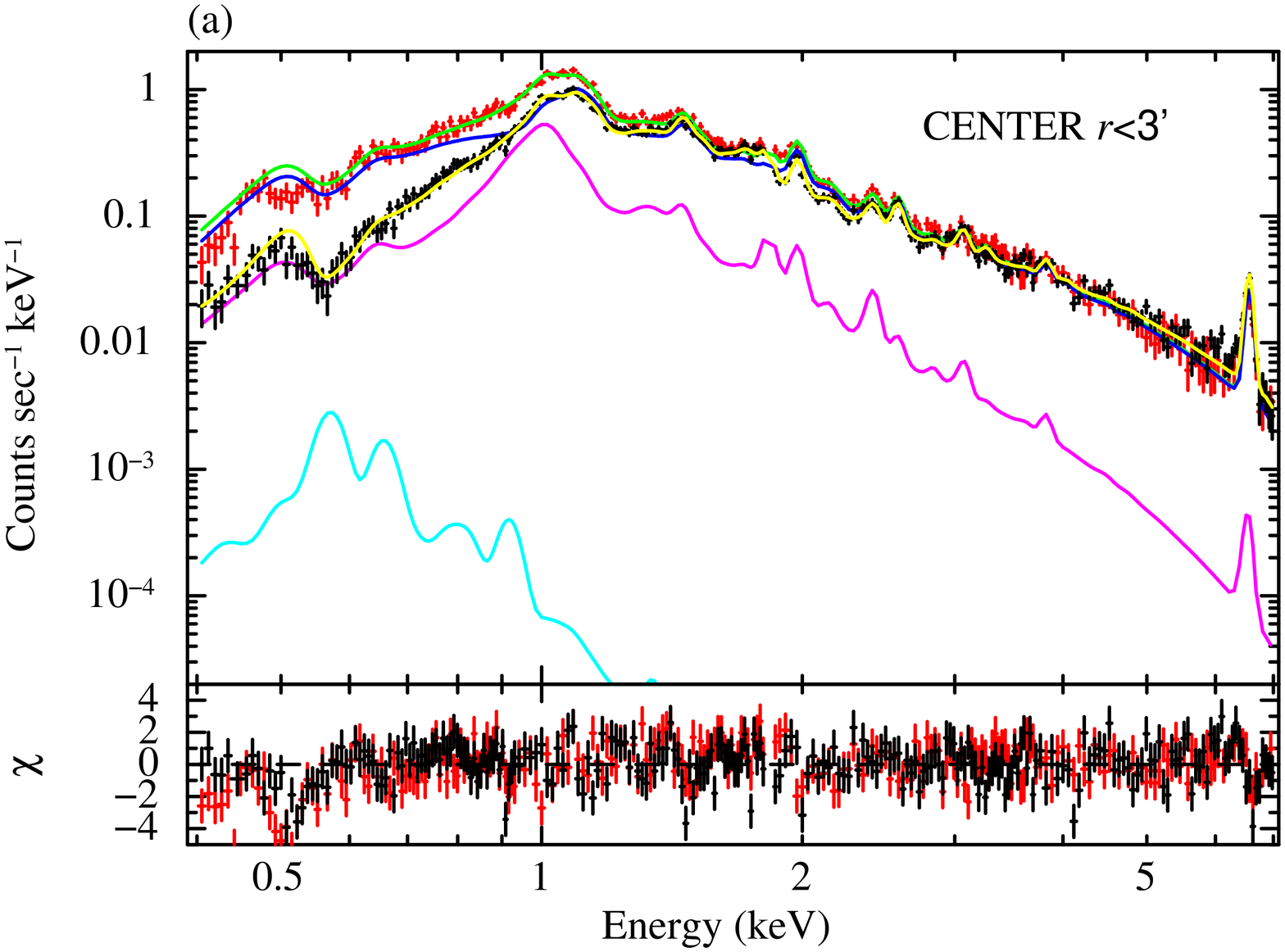}
\end{minipage}\hfill
\begin{minipage}{0.33\textwidth}
\FigureFile(\textwidth,\textwidth){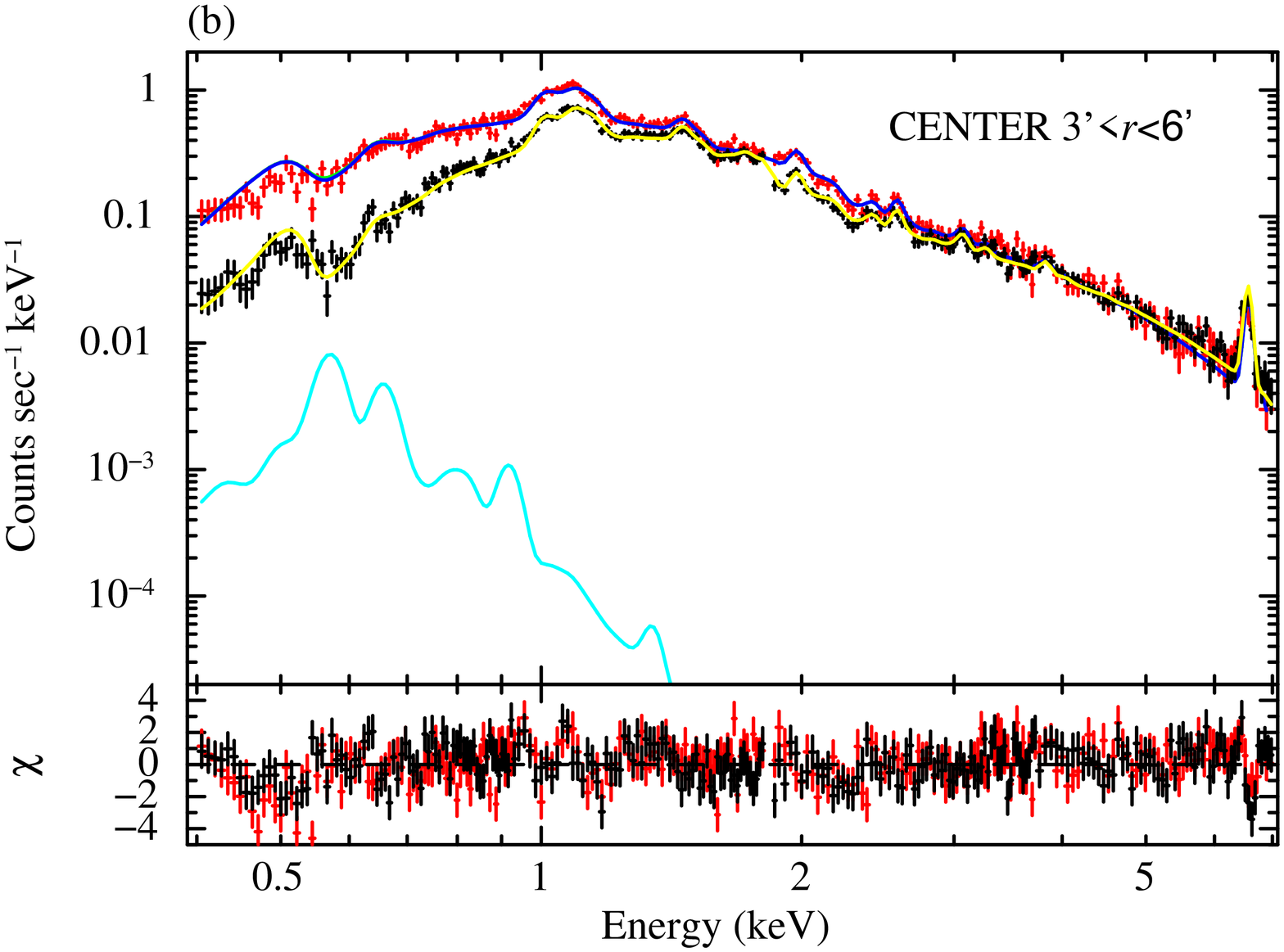}
\end{minipage}\hfill
\begin{minipage}{0.33\textwidth}
\FigureFile(\textwidth,\textwidth){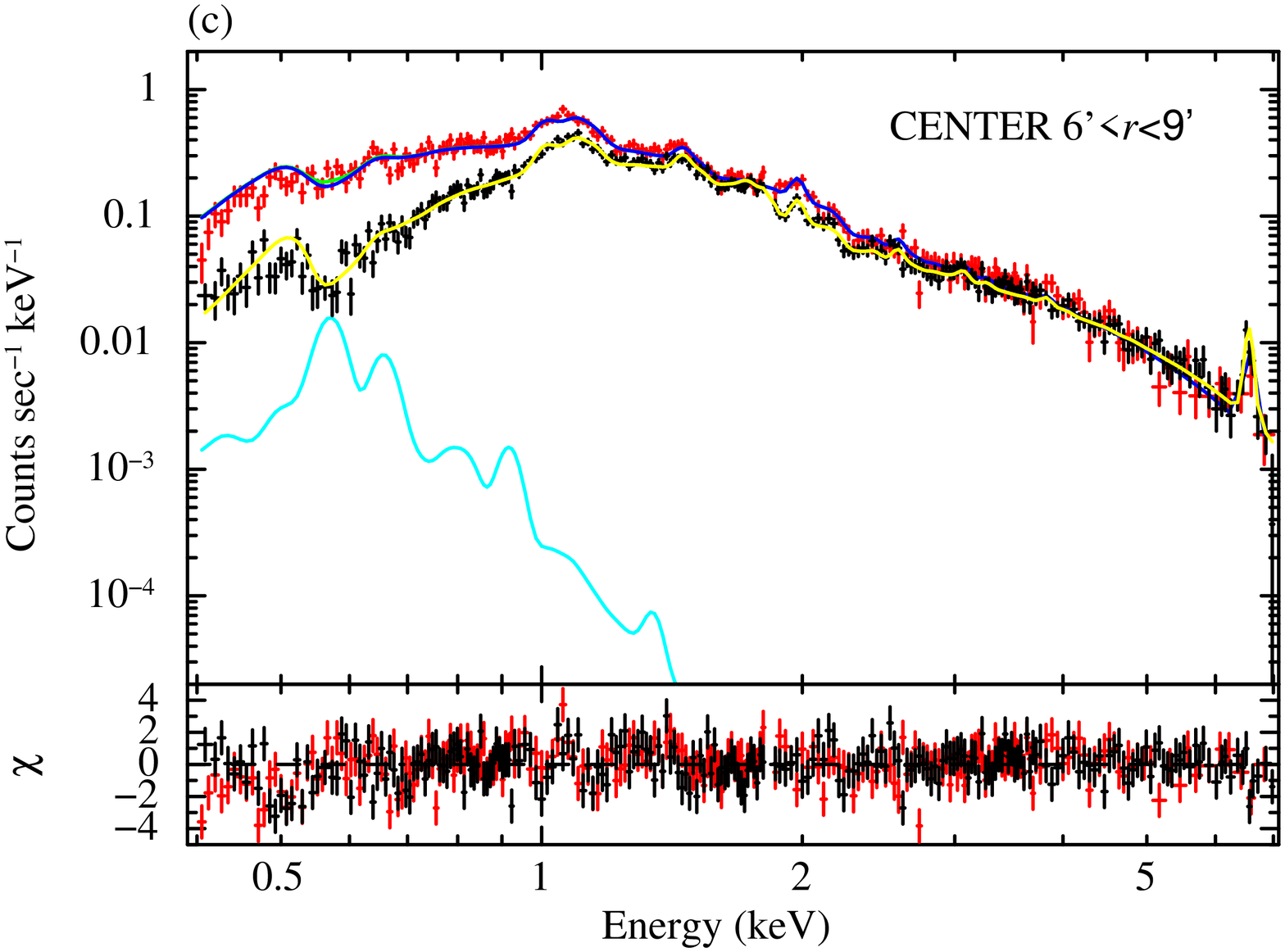}
\end{minipage}

\begin{minipage}{0.33\textwidth}
\FigureFile(\textwidth,\textwidth){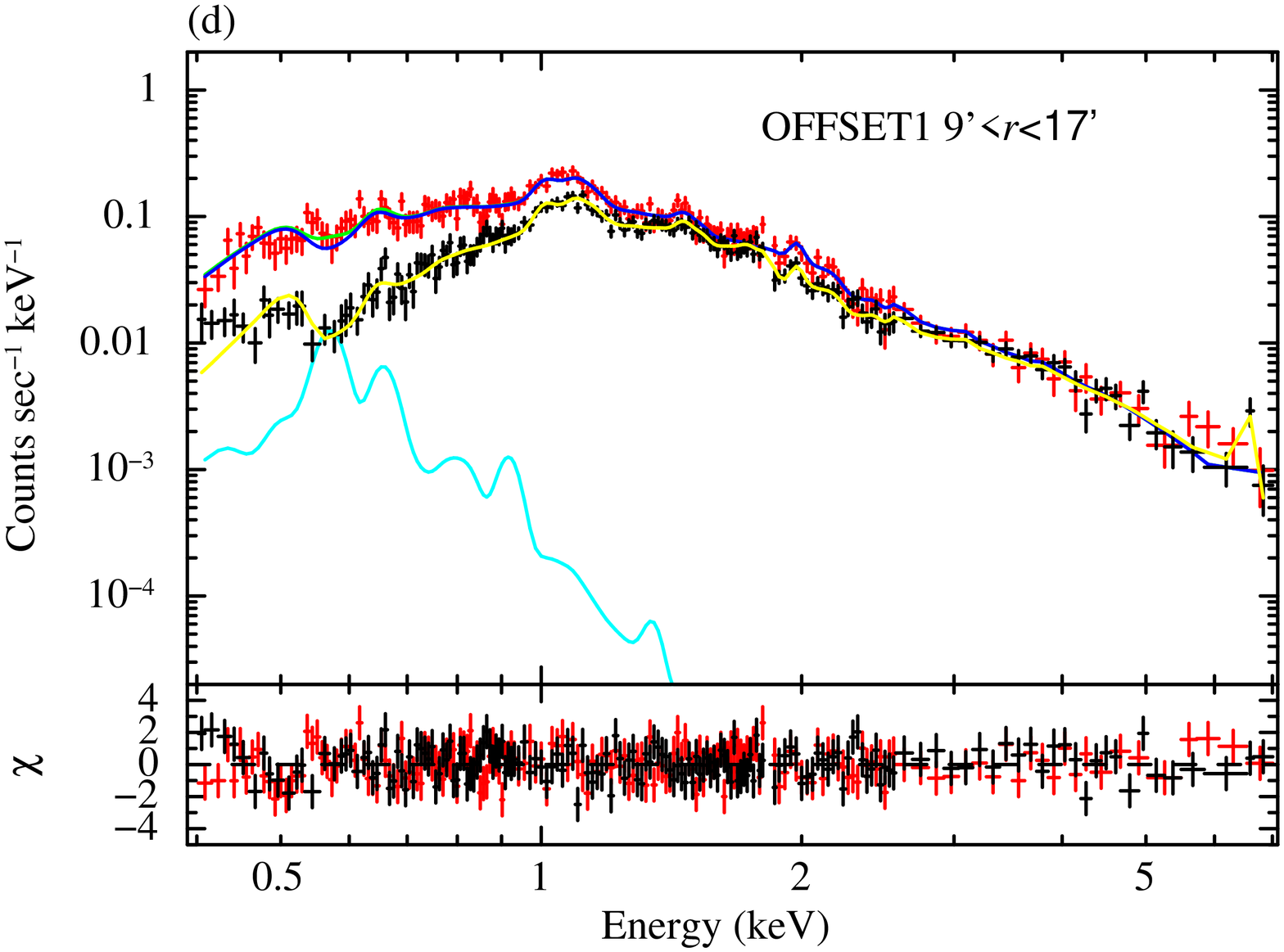}
\end{minipage}\hfill
\begin{minipage}{0.33\textwidth}
\FigureFile(\textwidth,\textwidth){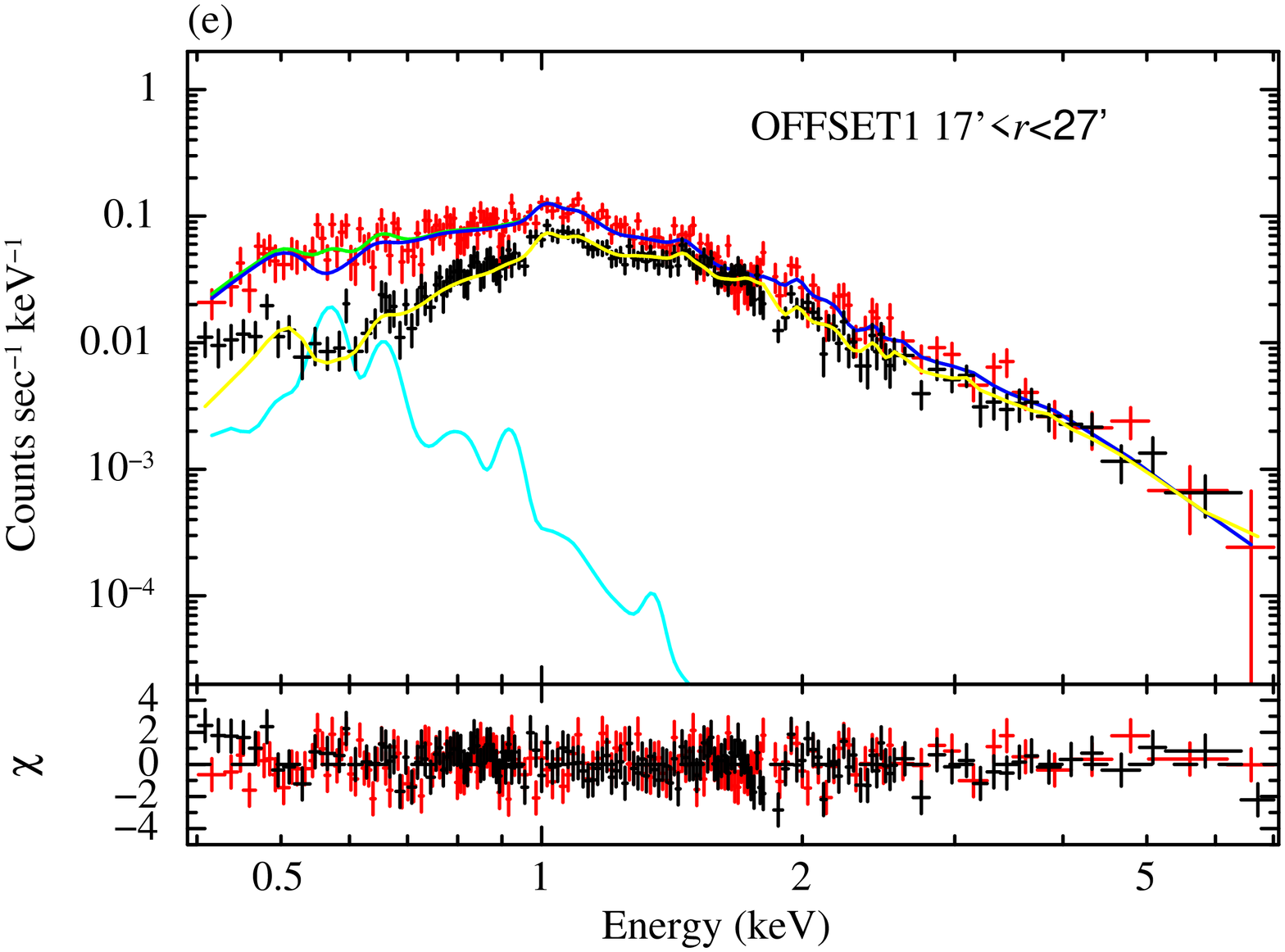}
\end{minipage}\hfill
\begin{minipage}{0.33\textwidth}
\FigureFile(\textwidth,\textwidth){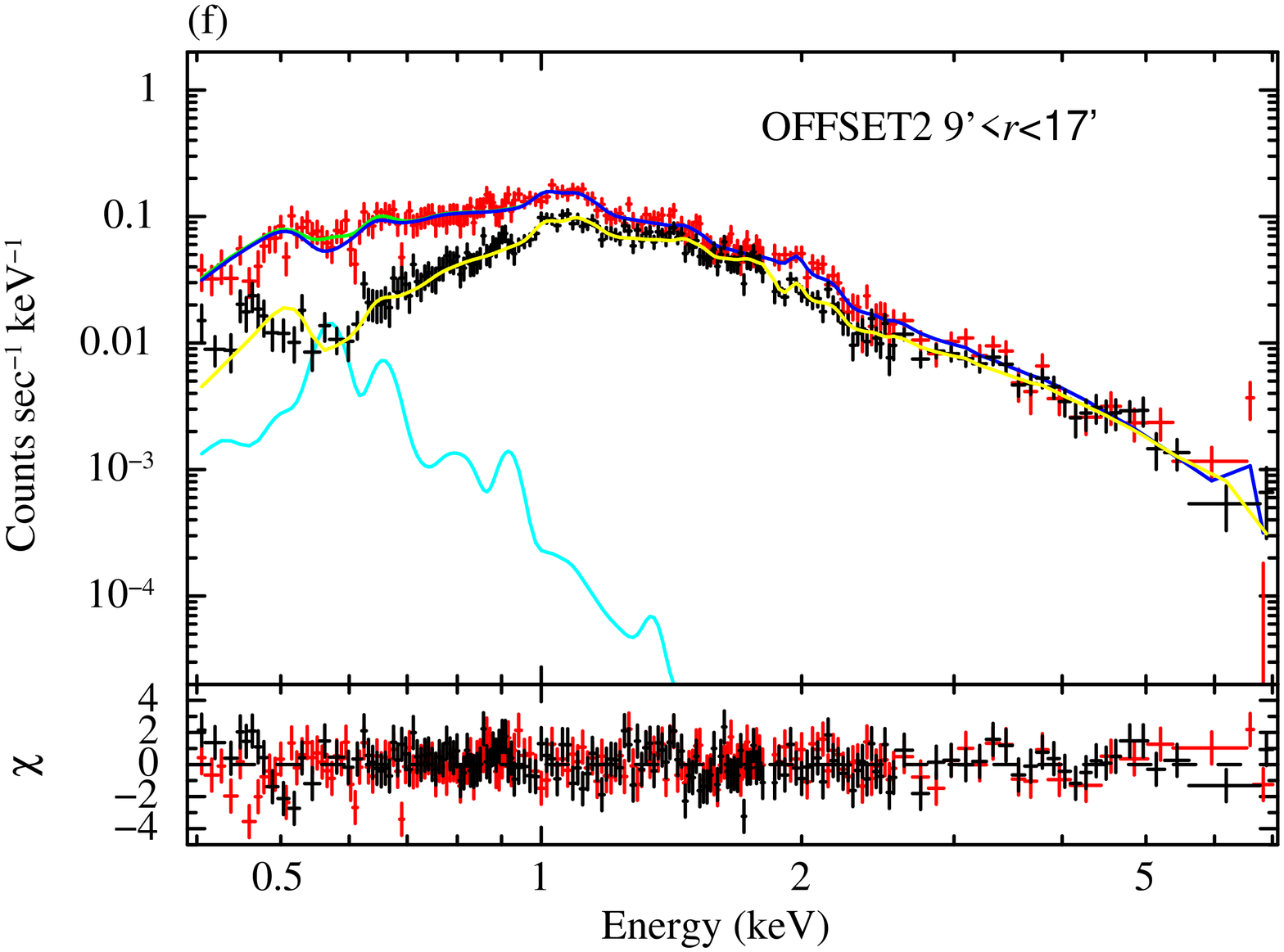}
\end{minipage}

\begin{minipage}{0.33\textwidth}
\FigureFile(\textwidth,\textwidth){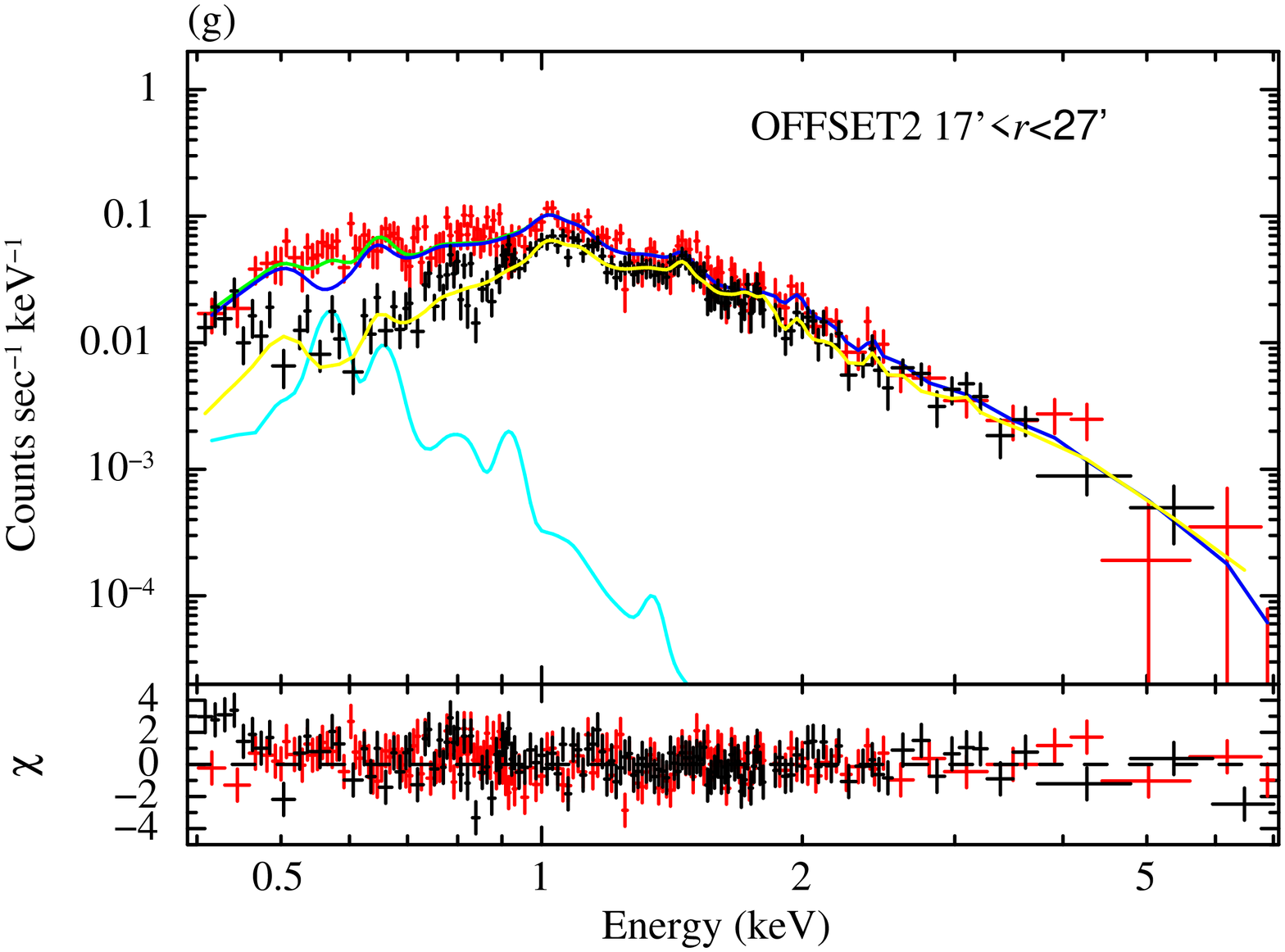}
\end{minipage}\hfill
\begin{minipage}{0.66\textwidth}
\caption{The panels show the observed spectra
for the annular regions of A~262 which are denoted in the panels,
and the data are plotted by red and black crosses for BI and FI, respectively.
The estimated CXB and NXB components are subtracted,
and the green and yellow lines show the best-fit model 
for the BI and FI spectra, respectively.
The BI spectra of the ICM component are shown in blue and 
magenta lines, and the Galactic emission 
is indicated by cyan line.
The energy range around the Si K-edge (1.825--1.840 keV) is ignored
in the spectral fits.
The lower panels show the fit residuals in units of $\sigma$.
}\label{fig:2}
\end{minipage}
\end{figure*}

We extracted spectra from five annular regions
of 0$'$--3$'$, 3$'$--6$'$, and 6$'$--9$'$ for the central observation, 
and 9$'$--17$'$, 17$'$--27$'$ for the offset regions,
centered on (RA, Dec) = (\timeform{1h52m46.1s}, \timeform{+36D09'33''}).
Table~\ref{tab:2} lists the areas of the extracted regions (arcmin$^2$),
fractional coverage of the annulus (\%),
the {\sc source\_ratio\_reg} values (\%; see caption for its definition)
and the BI and FI counts for the observed spectra and the estimated 
NXB and CXB spectra. The fraction of the background,
$f_{\rm BGD}\equiv\rm (NXB + CXB)/OBS$, was less than $\sim$50\%
even at the outermost annulus, although the Galactic component
is not considered here.
Each annular spectrum is shown in figure~\ref{fig:2}.
The ionized Mg, Si, S, Fe lines are clearly seen in each region.
The O\emissiontype{VII} and O\emissiontype{VIII} lines were prominent
in the outer rings, however, most of the O\emissiontype{VII} line
was considered to come from the local Galactic emission,
and we dealt with those in the same way as \citet{sato07a} and 
\citet{sato08a}.

\begin{figure*}
\begin{minipage}{0.33\textwidth}
\FigureFile(\textwidth,\textwidth){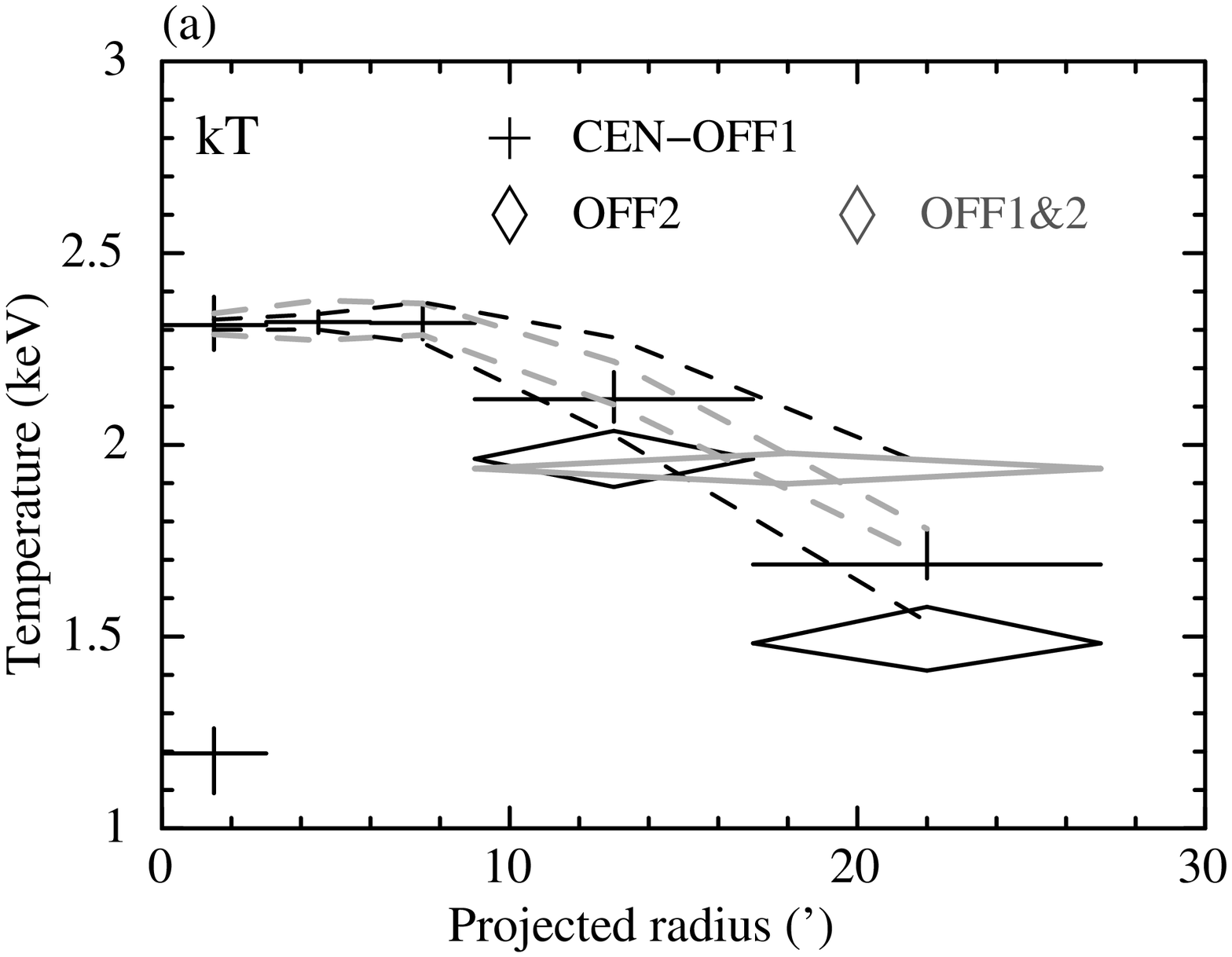}
\end{minipage}\hfill
\begin{minipage}{0.33\textwidth}
\FigureFile(\textwidth,\textwidth){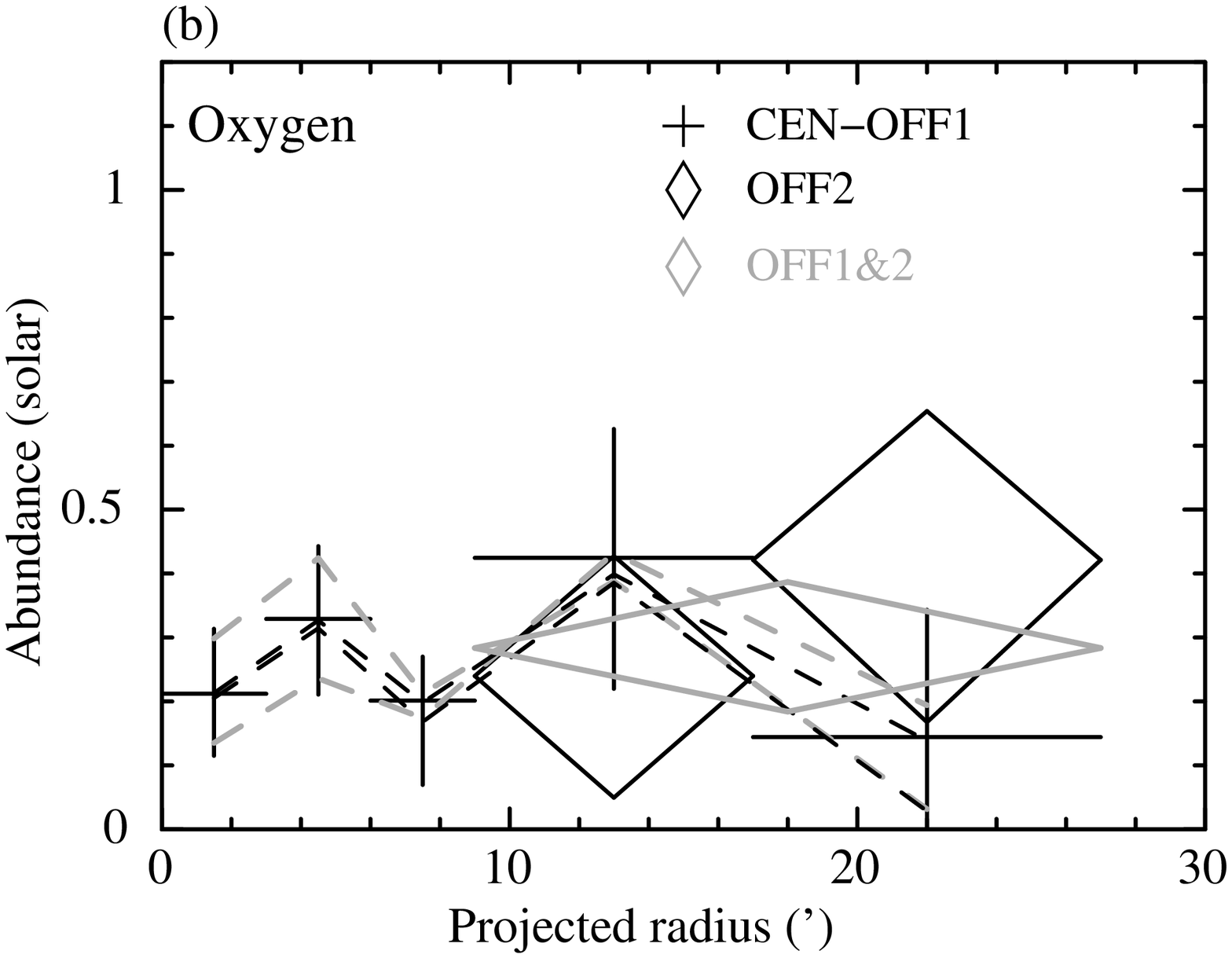}
\end{minipage}\hfill
\begin{minipage}{0.33\textwidth}
\FigureFile(\textwidth,\textwidth){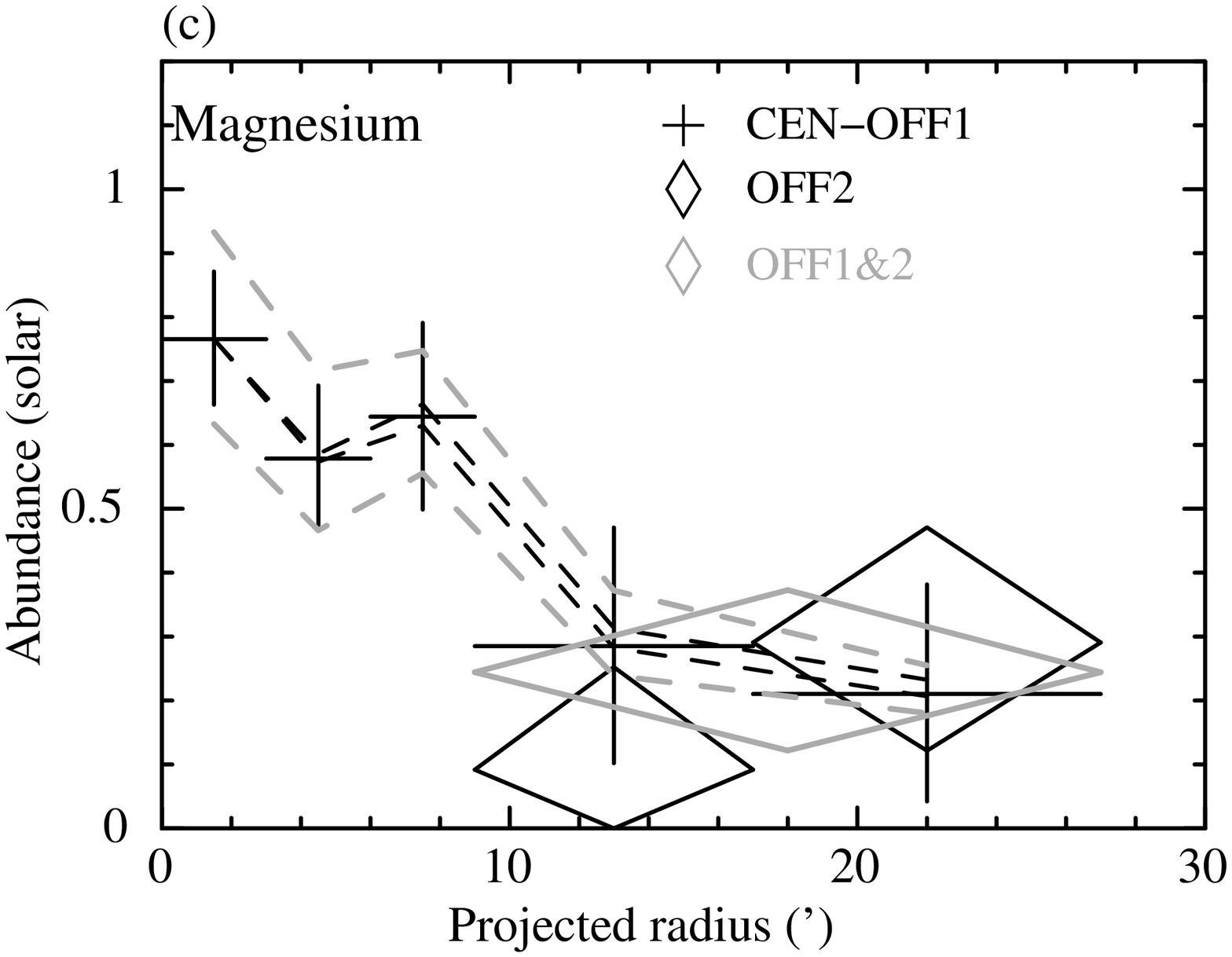}
\end{minipage}

\begin{minipage}{0.33\textwidth}
\FigureFile(\textwidth,\textwidth){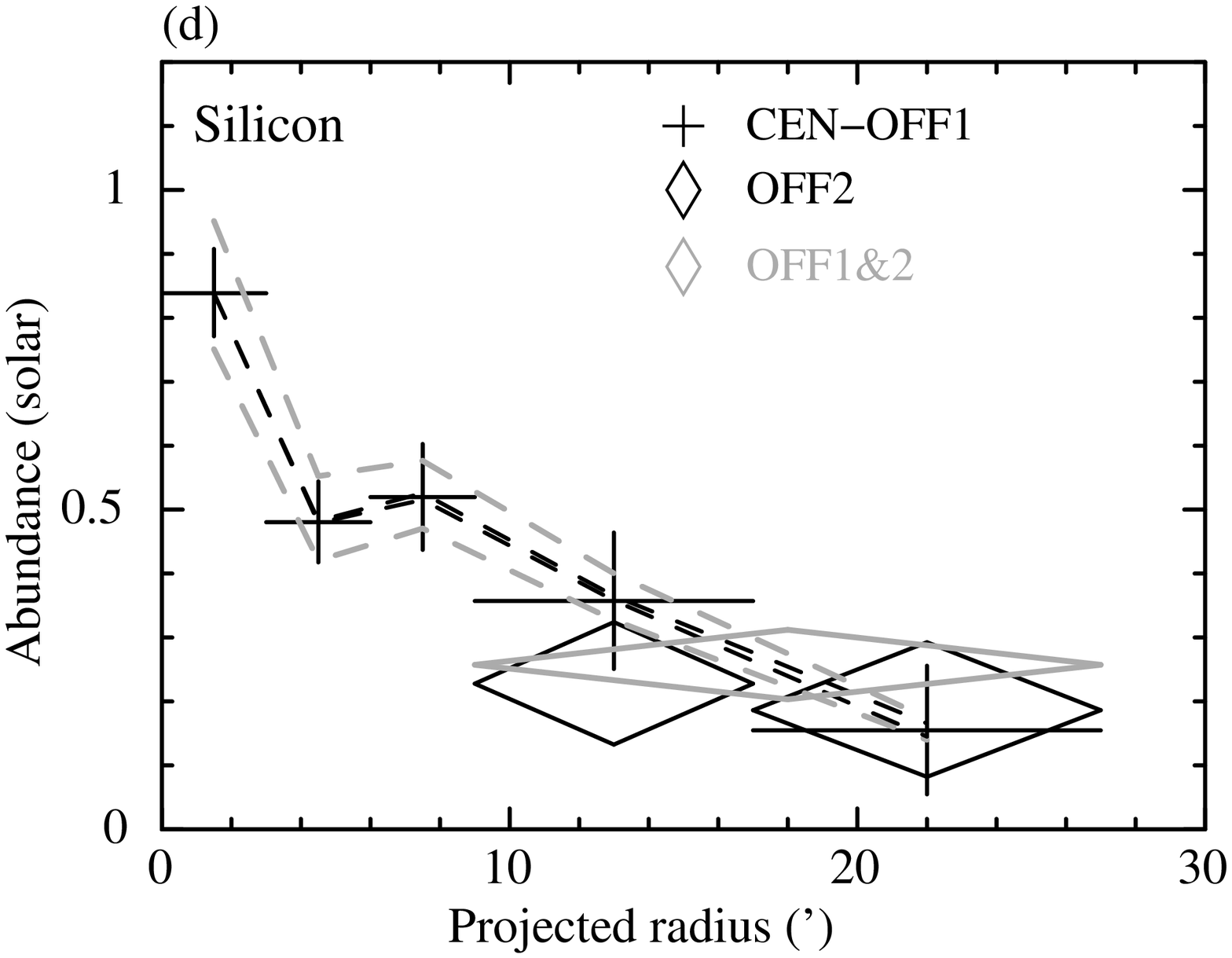}
\end{minipage}\hfill
\begin{minipage}{0.33\textwidth}
\FigureFile(\textwidth,\textwidth){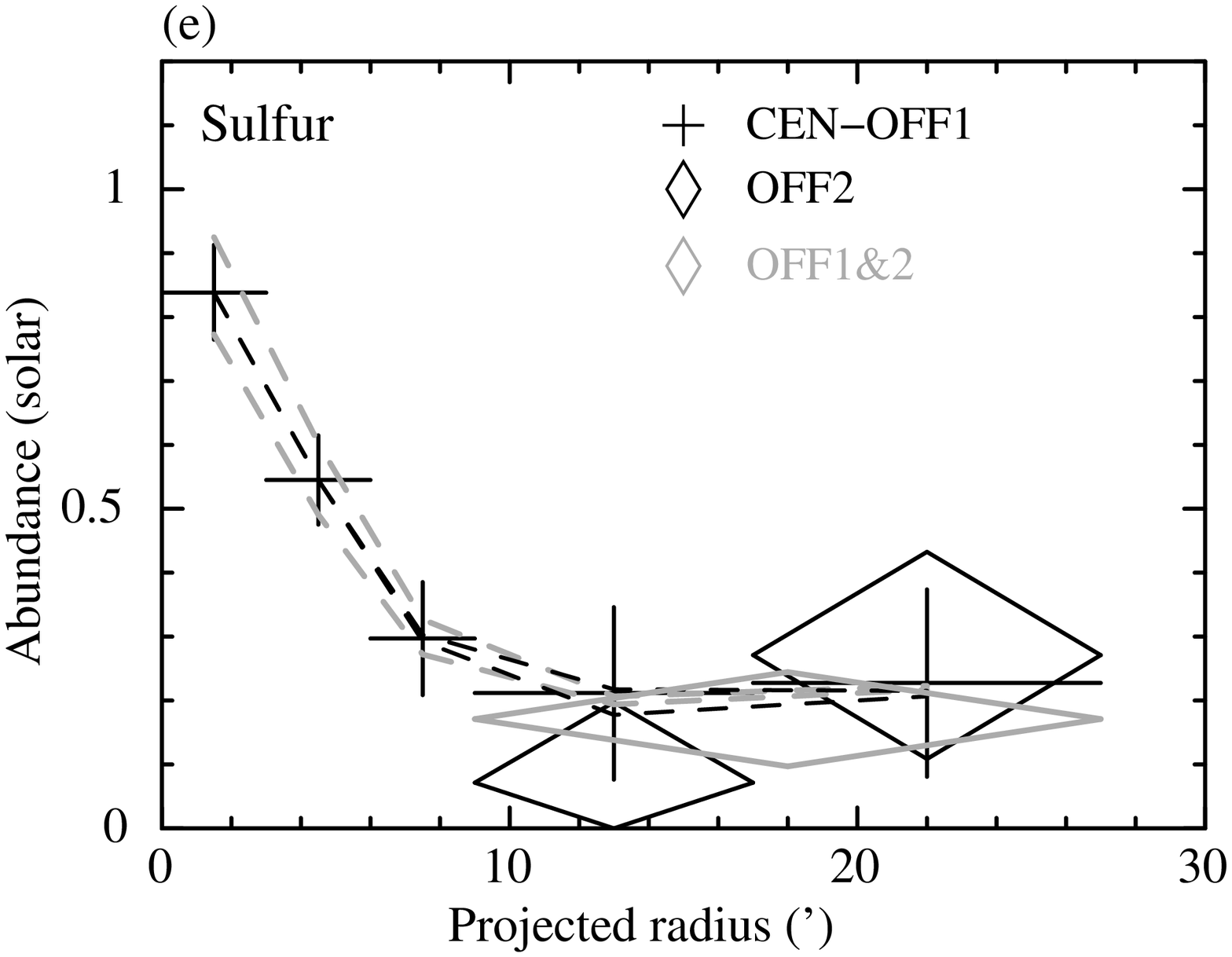}
\end{minipage}\hfill
\begin{minipage}{0.33\textwidth}
\FigureFile(\textwidth,\textwidth){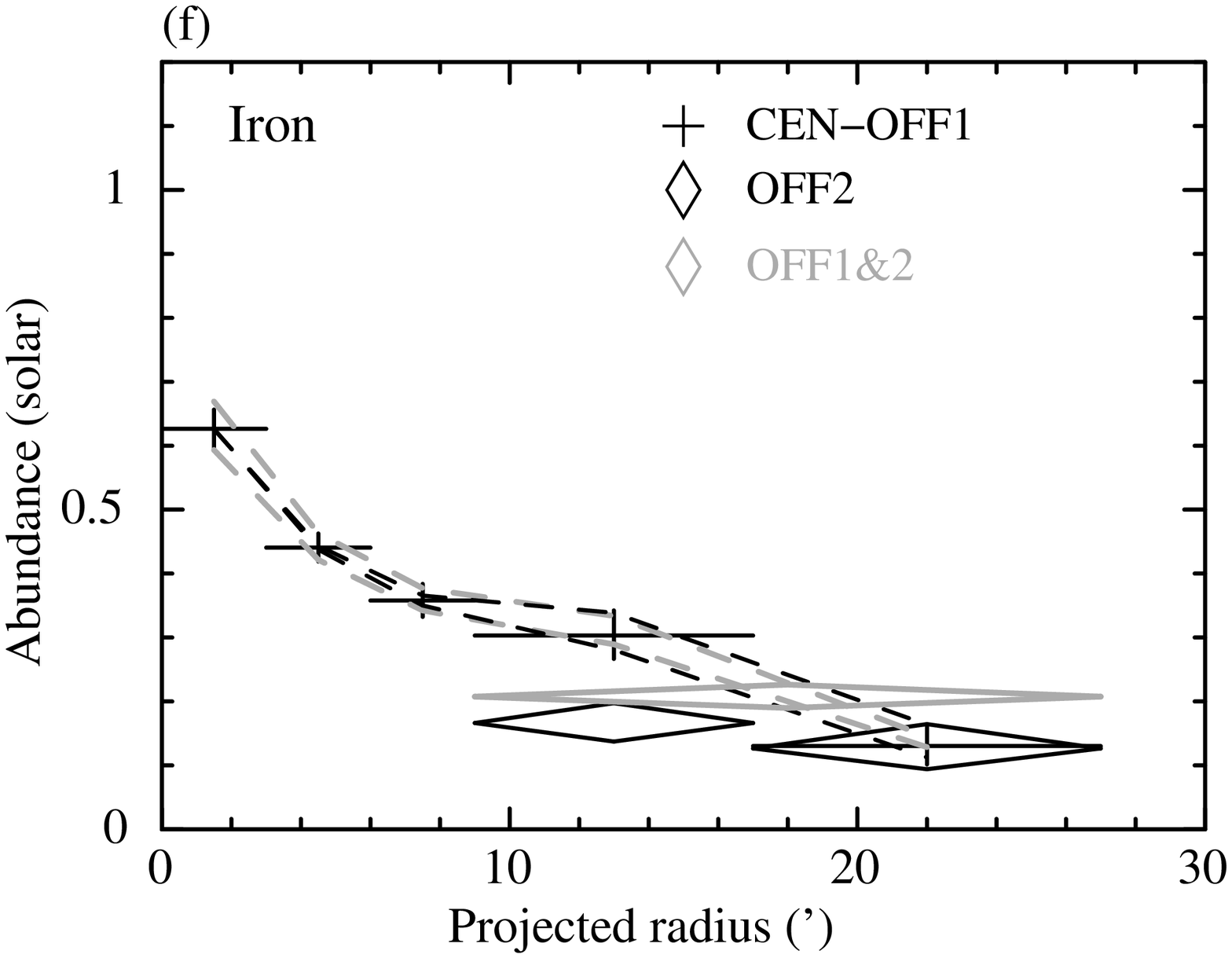}
\end{minipage}

\vspace*{-0.5ex}
\caption{(a): Radial temperature profiles derived from the spectral
fits for each annulus against the projected radius.  Light-gray
diamonds correspond to the results of the addition of offset 1 and 2
spectra.  Black dashed lines show systematic change of the best-fit
values of the center--offset1 region by varying the thickness of the
OBF contaminant by $\pm 10$\%.  Light-gray dashed lines denote those
when the estimated CXB and NXB levels are varied by $\pm 10$\%.
(b)--(f): Radial abundance profiles are plotted in the same
way as in (a).}
\label{fig:3}
\end{figure*}

\begin{table*}
\caption{Summary of the parameters of the fits to
each annular spectrum of A~262.  All annuli were simultaneously
fitted.  The values of Offset 1\&2 correspond to the results of the 
summed spectra in the offset 1 and 2 regions. 
Errors are 90\% confidence range of statistical errors, and
do not include systematic errors.  The solar abundance ratio of {\it
angr} was assumed.  These results are plotted in figure~\ref{fig:3}.
}\label{tab:3}
\begin{center}
\begin{tabular}{lcccccc}
\hline\hline
\makebox[2em][l]{Galactic} & {\it Norm}$_1$\makebox[0in][l]{\,$^\ast$} & $kT_1$\makebox[0in][l]{\,$^\dagger$}& {\it Norm}$_2$\makebox[0in][l]{\,$^\ast$} & $kT_2$&  &  \\
& &(keV)&&(keV)& & \\
\hline
     & 0.00$^{+2.25}_{-0.00}$ & 0.074(fix) & 0.96$^{+0.54}_{-0.28}$ & 0.174$^{+0.036}_{-0.043}$ &  &  \\
\hline\hline
\makebox[2em][l]{ICM} & {\it Norm}$_1$\makebox[0in][l]{\,$^\ddagger$} & $kT_1$& {\it Norm}$_2$\makebox[0in][l]{\,$^\ddagger$} & $kT_2$& {\it Norm}$_1$/{\it Norm}$_2$ & $\chi^2$/dof\makebox[0in][l]{\,$^\S$} \\
& &(keV)&&(keV)& & \\
\hline
Center & & & & & & \\
0$'$--3$'$    &  365.4$^{+19.9}_{-20.7}$  & 2.31$^{+0.07}_{-0.06}$ &  67.0$^{+21.0}_{-22.3}$  & 1.20$^{+0.07}_{-0.10}$ & 5.41$^{+1.71}_{-1.81}$ &  895/497\\
3$'$--6$'$    &  150.7$^{+3.1}_{-2.8}$  & 2.32$^{+0.03}_{-0.03}$ & -- & -- & -- & 726/498\\
6$'$--9$'$    & 71.7$^{+1.8}_{-1.7}$  & 2.32$^{+0.04}_{-0.04}$ & -- & -- & -- & 648/498\\
\hline
Offset1 & & & & & & \\
9$'$--17$'$    &  5.2$^{+0.2}_{-0.2}$  & 2.12$^{+0.07}_{-0.06}$ & -- & -- & -- & 364/382\\
17$'$--27$'$    & 1.7$^{+0.1}_{-0.1}$  & 1.69$^{+0.09}_{-0.04}$ & -- & -- & -- & 399/382\\
\hline
Offset2 & & & & & & \\
9$'$--17$'$    & 5.2$^{+0.2}_{-0.2}$  & 1.96$^{+0.07}_{-0.07}$ & -- & -- & -- & 408/382\\
17$'$--27$'$    &  1.2$^{+0.1}_{-0.1}$ & 1.48$^{+0.09}_{-0.07}$ & -- & -- & -- & 530/379\\
total  &  &  &  &  &  & 3970/3018\\\\[-2ex]
\hline
Offset1\&2 & & & & & & \\
9$'$--27$'$    & 8.6$^{+0.2}_{-0.2}$  & 1.94$^{+0.04}_{-0.04}$ & -- & -- & -- & 4149/3044\\
\hline\hline
\makebox[2em][l]{ICM} &O&Ne&Mg,Al&Si&\makebox[0in][c]{S,Ar,Ca}&Fe,Ni \\
&(solar)&(solar)&(solar)&(solar)&(solar)&(solar) \\
\hline
Center & & & & & & \\
0$'$--3$'$    & 0.21$^{+0.10}_{-0.10}$ & 0.65$^{+0.21}_{-0.20}$ & 0.77$^{+0.11}_{-0.10}$ & 0.84$^{+0.07}_{-0.07}$ & 0.84$^{+0.07}_{-0.07}$ & 0.63$^{+0.03}_{-0.03}$ \\
3$'$--6$'$    & 0.33$^{+0.11}_{-0.12}$ & 1.08$^{+0.15}_{-0.15}$ & 0.58$^{+0.11}_{-0.11}$ & 0.48$^{+0.06}_{-0.06}$ & 0.55$^{+0.07}_{-0.07}$ & 0.44$^{+0.02}_{-0.02}$ \\
6$'$--9$'$    & 0.20$^{+0.07}_{-0.13}$ & 0.88$^{+0.18}_{-0.17}$ & 0.64$^{+0.15}_{-0.15}$ & 0.52$^{+0.08}_{-0.08}$  & 0.30$^{+0.09}_{-0.09}$ & 0.36$^{+0.03}_{-0.03}$ \\
\hline
Offset1 & & & & & & \\
9$'$--17$'$    & 0.42$^{+0.20}_{-0.20}$ & 0.77$^{+0.28}_{-0.26}$ & 0.29$^{+0.19}_{-0.18}$ & 0.36$^{+0.11}_{-0.11}$ & 0.21$^{+0.13}_{-0.14}$ & 0.30$^{+0.04}_{-0.04}$ \\
17$'$--27$'$    & 0.14$^{+0.20}_{-0.14}$ & 0.42$^{+0.27}_{-0.14}$ & 0.21$^{+0.17}_{-0.17}$ & 0.15$^{+0.10}_{-0.10}$ & 0.23$^{+0.15}_{-0.15}$ & 0.13$^{+0.03}_{-0.03}$ \\
\hline
Offset2 & & & & & & \\
9$'$--17$'$    & 0.24$^{+0.19}_{-0.19}$ & 0.34$^{+0.21}_{-0.20}$ & 0.09$^{+0.16}_{-0.09}$ & 0.23$^{+0.10}_{-0.10}$ & 0.07$^{+0.13}_{-0.07}$ & 0.17$^{+0.03}_{-0.03}$ \\
17$'$--27$'$    & 0.42$^{+0.23}_{-0.25}$ & 0.19$^{+0.26}_{-0.19}$ & 0.29$^{+0.18}_{-0.17}$ & 0.19$^{+0.11}_{-0.10}$ & 0.27$^{+0.16}_{-0.16}$ & 0.13$^{+0.04}_{-0.03}$ \\
\hline
Offset1\&2 & & & & & & \\
9$'$--27$'$    & 0.28$^{+0.10}_{-0.10}$ & 0.54$^{+0.13}_{-0.12}$ & 0.24$^{+0.09}_{-0.09}$ & 0.26$^{+0.05}_{-0.05}$ & 0.17$^{+0.07}_{-0.07}$ & 0.21$^{+0.02}_{-0.02}$ \\
\hline\\[-1ex]
\multicolumn{7}{l}{\parbox{0.75\textwidth}{\footnotesize 
\footnotemark[$\ast$] 
Normalization of the {\it apec} component
divided by the solid angle, $\Omega^{\makebox{\tiny\sc u}}$,
assumed in the uniform-sky ARF calculation (20$'$ radius),
${\it Norm} = \int n_{\rm e} n_{\rm H} dV \,/\,
(4\pi\, (1+z)^2 D_{\rm A}^{\,2}) \,/\, \Omega^{\makebox{\tiny\sc u}}$
$\times 10^{-20}$ cm$^{-5}$~arcmin$^{-2}$, 
where $D_{\rm A}$ is the angular distance to the source.}}\\
\multicolumn{7}{l}{\parbox{0.75\textwidth}{\footnotesize
\footnotemark[$\dagger$] 
The value is shown in \citet{lumb02}.
}}\\
\multicolumn{7}{l}{\parbox{0.75\textwidth}{\footnotesize
\footnotemark[$\ddagger$] 
Normalization of the {\it vapec} component scaled with a factor of
{\sc source\_ratio\_reg} / {\sc area} in table~\ref{tab:2},\\ ${\it
Norm}=\frac{\makebox{\sc source\_ratio\_reg}}{\makebox{\sc area}} \int
n_{\rm e} n_{\rm H} dV \,/\, [4\pi\, (1+z)^2 D_{\rm A}^{\,2}]$ $\times
10^{-20}$~cm$^{-5}$~arcmin$^{-2}$, where $D_{\rm A}$ is the angular
distance to the source. }}\\
\multicolumn{7}{l}{\parbox{0.75\textwidth}{\footnotesize
\footnotemark[$\S$] 
All regions were fitted simultaneously.
}}
\end{tabular}
\end{center}
\end{table*}

The spectra with BI and FI for all regions were fitted simultaneously
in the energy range 0.4--7.1 keV\@.  In the simultaneous fit, the
common Galactic emission component was included for all regions, while
the A~262 emission was independently adjusted in each region.  We
excluded the narrow energy band around the Si K-edge
(1.825--1.840~keV) because its response was not modeled correctly.
The energy range below 0.4~keV was also excluded because the C edge
(0.284~keV) seen in the BI spectra could not be reproduced well in our
data.  The range above 7.1~keV was also ignored because Ni line ($\sim
7.5$~keV) in the background left a spurious feature after the NXB
subtraction at large radii.  In the simultaneous fits of BI and FI
data, only the normalization parameter was allowed to take different
values between them.

It is important to estimate the Galactic component precisely.  The
Galactic component gives significant contribution especially in the
outer regions, as shown in figure~\ref{fig:2}.  However, the ICM
component is still dominant in almost all the energy range except for
the O\emissiontype{VII} line.  We assumed two temperature {\it apec}
model (assuming 1 solar abundance with redshift $z=0$) for the
Galactic component, and fitted the data with the following model
formula: ${\it constant}\times( {\it apec}_{\rm cool}+{\it apec}_{\it
hot} + {\rm ICM components} )$, where we fixed the temperature of
${\it apec}_{\rm cool}$ to be 0.074 keV as given by in \citet{lumb02}.
As a result, the best-fit normalization of the cool (0.074 keV)
component turned out to be 0.  When we fix the two temperatures to be
the value in \citet{lumb02}, the resultant parameters of the fit 
did not change significantly.  Thus, we concluded that the
two temperature model of the Galactic emission (the cool component
temperature was fixed to be 0.074 keV) was enough to fit the A~262
data since the intensity of the Local Hot bubble in the A~262
direction seemed weak as shown in \citet{mccammon02}.  In addition, we
also examined the following model formula: ${\it constant}\times({\it
apec}_{\it hot} + {\it phabs} \times ({\it apec}_{\rm cool}+ {\rm ICM~
component})$, however the $\chi^2$ values with this model were only a
little worse than that with the above model.  With this latter model,
the results for the ICM component did not change significantly.  The
resultant normalizations of the {\it apec} models in table~\ref{tab:3}
are scaled so that they give the surface brightness in unit solid
angle of arcmin$^2$, and are constrained to give the same surface
brightness and the same temperature for the simultaneous fits of all
annuli.

The ICM spectrum for the central region, $r<3'$, was clearly better
represented by two {\it vapec} models than one {\it vapec} model in
the $\chi^2$ test. On the other hand, the ICM spectra for the outer
regions, $r>3'$, were well-presented by single temperature model.
Thus, we carried out the simultaneous fit with the following formula
of the Galactic and ICM components: ${\it constant}\times( {\it
apec}_{\rm cool}+{\it apec}_{\rm hot} + {\it phabs} \times ({\it
vapec}_{0<r<27'} + {\it vapec}_{0<r<3'})$\@.  The fit results are
shown in table \ref{tab:3}.  The abundances were linked in the
following way, Mg=Al, S=Ar=Ca, Fe=Ni, which gave good constraint
especially for the offset regions.  The abundances were also linked
between the two ${\it vapec}$ components for $r<3'$ region.  Results
of the spectral fit for individual annuli are summarized in
table~\ref{tab:3} and figure~\ref{fig:3}, in which systematic error
due to the OBF contamination and background (CXB + NXB) estimation are
shown.

\begin{table*}
\caption{List of $\chi^2$/dof for the fits of the nominal and
considering the systematic errors such as contaminant of OBF and 
background level. For details, see text.}
\label{tab:4}
\begin{center}
\begin{tabular}{lccccc}
\hline\hline
\makebox[6em][l]{Region} & nominal &\multicolumn{2}{c}{contaminant} & \multicolumn{2}{c}{background}\\
\hline
 & & \makebox[0in][c]{+10\%} & \makebox[0in][c]{-10\%} & \makebox[0in][c]{+10\%} & \makebox[0in][c]{-10\%}\\
\hline
All $\dotfill$   & 3970/3018 & 3769/3018 & 4350/3018 & 4041/3018 & 3997/3018 \\
\hline
\end{tabular}
\end{center}
\end{table*}

\subsection{Temperature Profile}
\label{subsec:radial}

Radial temperature profile and the ratio of the {\it vapec}
normalizations between the hot and cool ICM components are shown in
figure~\ref{fig:3}(a) and table~\ref{tab:3}. 
The ICM temperature of hot and cool components at the central region 
was $\sim2.3$ and $\sim1.2$ keV, respectively, and the temperature 
decreased mildly to $\sim1.5$ keV in the outermost region.
Our results for the two temperature ICM model are consistent with the
Chandra result \citep{blanton04} for the central region within $3'$.
For the outer region ($r>3'$), our results are also consistent with 
the previous Chandra result \citep{vikhlinin06}, and the 
Chandra/XMM results \citep{gastaldello07}. 
The radius of $27'\sim 540$~kpc corresponds to $\sim 0.43\; r_{\rm
180}$, and the temperature decline, observed in several other clusters, 
is clearly recognized in this system out to this radius.

\subsection{Abundance Profiles}

Metal abundances are determined for the six element groups
individually as shown in figures~\ref{fig:3}(b)--(f).  The four
abundance values for Mg, Si, S, and Fe and their radial variation look
quite similar to each other.  The central abundances lie around $\sim
1.0$ solar, and they commonly decline to about 1/5 of the central
value in the outermost annulus.  Note that, although the Fe abundance
decreased to $r>\sim20';~0.3~r_{180}$, the abundance at the outermost
region was determined with Fe-L line as shown in figure~\ref{fig:2}
because of the poor statistics around Fe-K line.  On the other hand,
the O profile looks flatter compared with the other elements.  Because
the results for the offset regions had large errors, we examined the
summed spectra for these regions.  However, even if the spectra in the
$9'<r<17'$ and $17'<27'$ regions for both the offset1 and 2
observations were combined, the tendency of the radial abundance
distributions did not change significantly. 
We also noted that, when we examined all abundances to be free in the fits,
the resultant parameters did not change within the statistical errors.

\subsection{Systematic Errors and Uncertainties}

We performed the spectral fits with the {\it vmekal} model instead of 
the {\it vapec} model. As a result, the abundance profiles did not
change within the statistical errors, although the Mg abundances 
with the {\it vmekal} model were $\sim30$--40\% lower than those 
with the {\it vapec} model.

We examined the systematic error of our results by changing the
background normalization by $\pm 10$\%, and the error range is plotted
with light-gray dashed lines in figure~\ref{fig:3}. The systematic
error due to the background estimation is almost negligible.  The
other systematic error concerning the uncertainty in the OBF
contaminant is shown by black dashed lines as shown in
figure~\ref{fig:3}.  A list of $\chi^2$/dof by changing the systematic
errors is presented in table \ref{tab:4}.  Although the fit is not
statistically acceptable due mainly to the very high photon statistics
compared with the systematic errors in the instrumental response,
these results are useful to assess whether each element abundance is
reasonably determined or not.  We note that Ne abundance is not
reliably determined due to an overlap with the strong and complex Fe-L
line emissions, however we left these abundance to vary freely during
the spectral fit.

\begin{figure*}
\centerline{
\FigureFile(0.45\textwidth,1cm){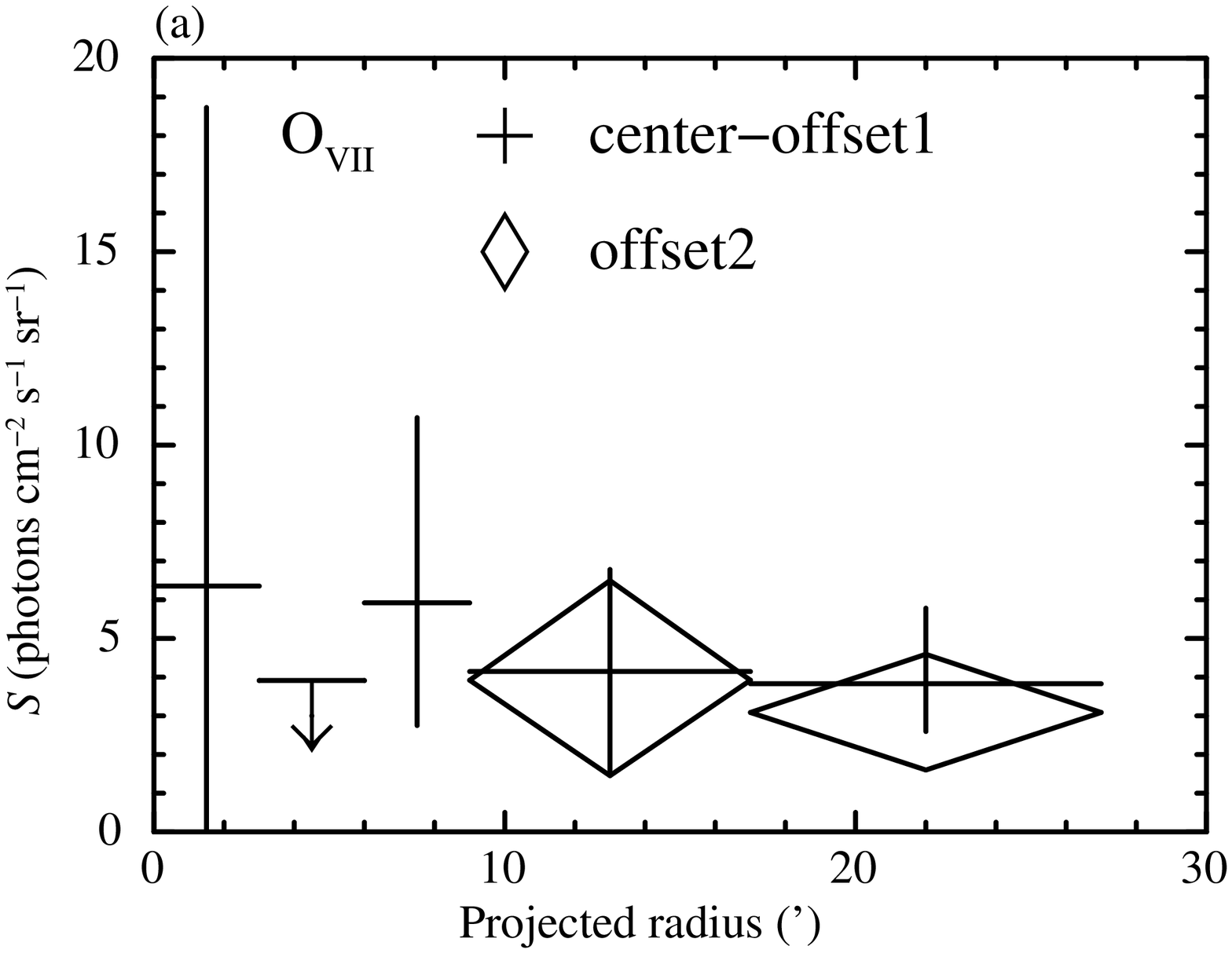}%
\hspace*{0.05\textwidth}
\FigureFile(0.45\textwidth,1cm){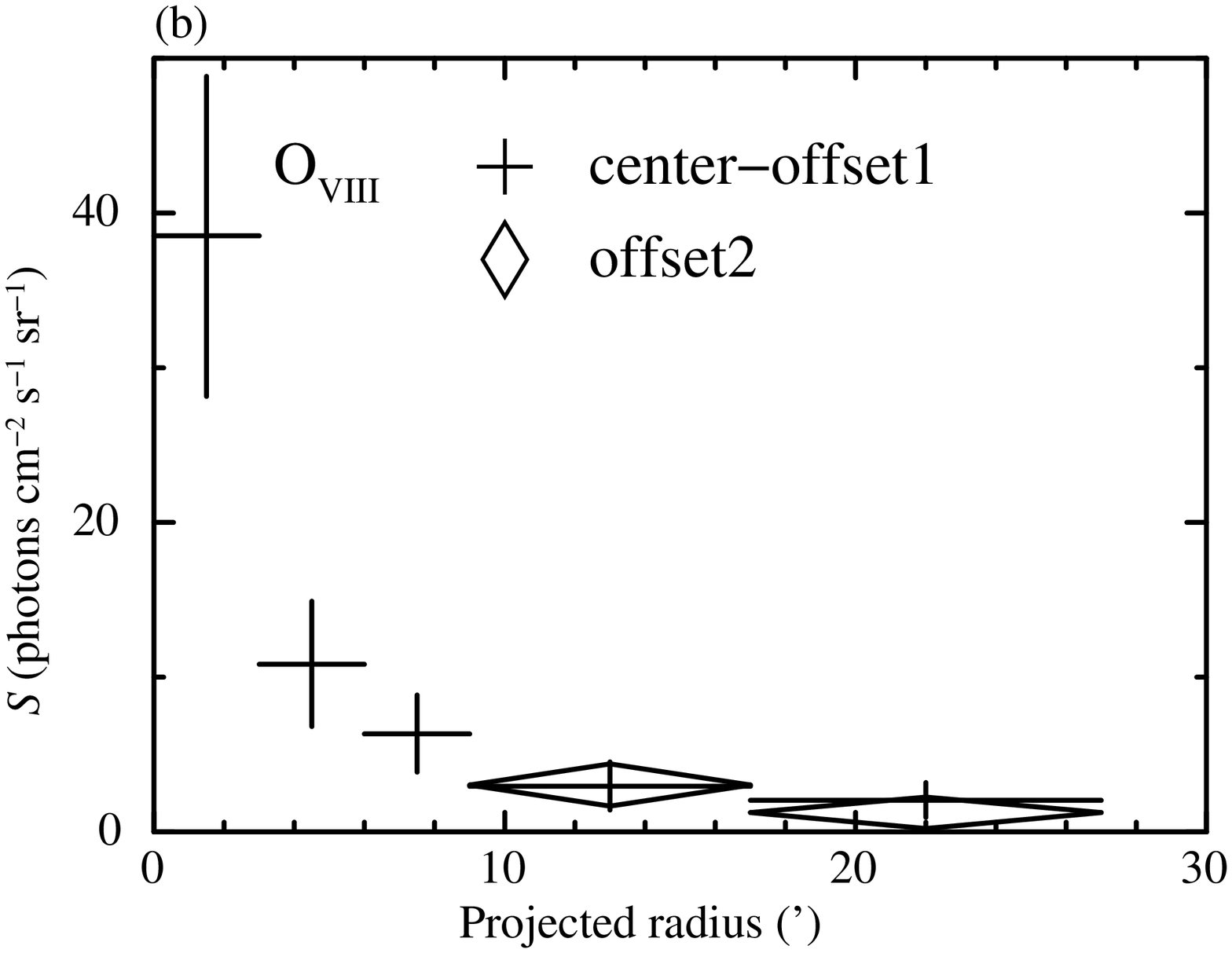}%
}
\caption{
Intensities of (a) O\emissiontype{VII} and (b) O\emissiontype{VIII} lines
at each annulus of A~262 in units of photons~cm$^{-2}$~s$^{-1}$~sr$^{-1}$.
}\label{fig:4}
\end{figure*}

Oxygen abundance is strongly affected by the Galactic emission as
shown in \citet{sato07a}.  However, we have shown that the use of
different temperature models for the Galactic emission did not
significantly affect the abundance results as mentioned in subsection
3.1\@.  We also examined radial surface brightness profiles of the
O\emissiontype{VII} and O\emissiontype{VIII} emission lines in order
to see if they are spatially uniform or peaked at the cluster
center. The surface brightness of the lines was derived by fitting the
annular spectrum with a {\it power-law} + {\it gaussian} + {\it
gaussian} model for an energy range between 500 and 700 eV\@. In this
fit, we fixed the Gaussian $\sigma$ to be 0, and allowed the energy
center of the two Gaussians to vary within 555--585~eV or 630--660~eV
for O\emissiontype{VII} or O\emissiontype{VIII} line, respectively.
The derived line intensities are shown in figure~\ref{fig:4}.  There
is a clear excess in the O\emissiontype{VIII} intensity profile
towards the cluster center, while O\emissiontype{VII} one is
consistent to be constant as already noticed by
\citet{sato07a,sato08a,sato08b}.  This is a clear evidence that the
O\emissiontype{VIII} line is associated with the ICM itself, on the
other hand, O\emissiontype{VII} line may mainly come from the Galaxy
or from the interplanetary space.

\section{Discussion}\label{sec:discuss}

\subsection{Metallicity Distribution in the ICM}

\begin{figure}
\centerline{\FigureFile(0.45\textwidth,8cm){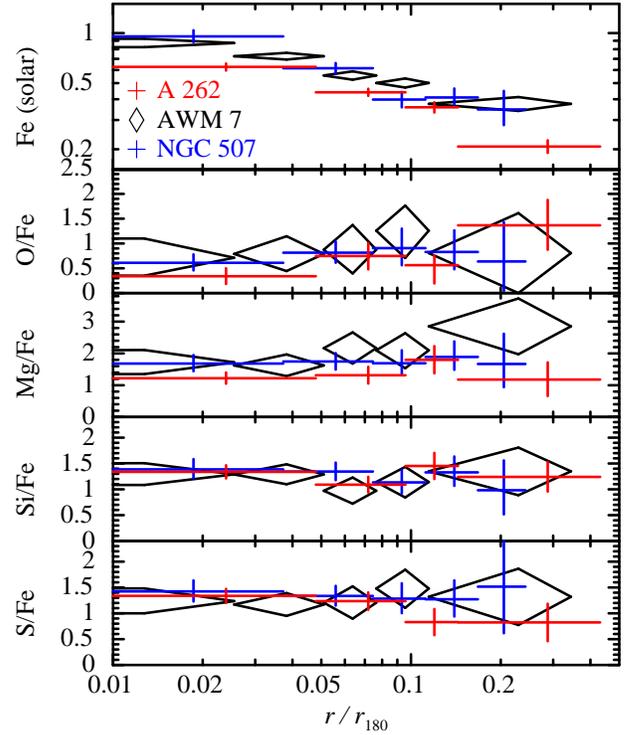}}
\vspace*{-1ex}
\caption{ Comparison of the abundance ratio for A~262 (red) with
results for the AWM~7 cluster (black diamonds; \cite{sato08a}) 
and NGC~507 group (blue crosses; \cite{sato08b}). 
Fe abundance and the O/Fe,
Mg/Fe, Si/Fe, and S/Fe abundance ratios in solar units
\citep{anders89} are plotted against the projected radius scaled by
the virial radius, $r_{180}$, in all panels.  }\label{fig:5}
\end{figure}

The present Suzaku observation of A~262 shows the abundance distribution
of O, Mg, Si, S, and Fe out to a radius of $27'\simeq 540$~kpc, which
corresponds to $\sim0.43~r_{180}$, as shown in figure \ref{fig:3}. The Ne
abundance shows a large ambiguity due to the strong coupling with Fe-L
lines.  Distributions of Mg, Si, S, and Fe are similar to each other,
while O profile in the outer region shows a large uncertainty.  We
plotted abundance ratios of O, Mg, Si, and S over Fe as a function of
the projected radius in figure~\ref{fig:5}.  The ratios Mg/Fe, Si/Fe
and S/Fe are consistent with a constant value around 1.5--2, while
O/Fe ratio for the innermost region ($r<2'$) is lower around 0.5\@.
In addition, the O/Fe ratio suggests some increase with radius.  The
Fe abundance at the outermost region has a large uncertainty compared
with that in the inner regions, because the value in the outer region
was determined by Fe-L line as shown in figure~\ref{fig:2}\@.  We note
that these abundance profiles are not deconvolved and averaged over
the line of sight.

Recent Suzaku observations have presented abundance profiles in
several other systems: an elliptical galaxy NGC~720 \citep{tawara08},
a group of galaxies HCG~62 \citep{tokoi08} and NGC~507
\citep{sato08b}, the Fornax cluster and NGC~1404
\citep{matsushita07b}, and a cluster of galaxies Abell~1060
\citep{sato07a} and AWM~7 \citep{sato08a}.  All systems show very
similar value of Si/Fe ratio, to be 1--1.5.  Mg/Fe ratio is slightly
higher in HCG~62, Abell ~1060 and AWM~7 than in other systems.  We
compare metal abundances of A~262 with those of AWM~7 and NGC~507 as
shown in figure~\ref{fig:5}.  Because AWM~7 ($\sim3.5$ keV) and
NGC~507 ($\sim1.5$ keV) host cD galaxies at the center and are similar
to A~262 in the X-ray morphology.  A~262 may connect groups of
galaxies and poor clusters regarding our understanding of the cluster
evolution.  Although the Fe abundance of A~262 is slightly lower than
that of the AWM~7 and NGC~507 in the central region, the Fe profile to
the outer region looks quite similar.  Abundance ratios of O/Fe,
Mg/Fe, Si/Fe, and S/Fe are also quite similar between the three
systems.  Therefore, the abundance ratios show closer similarity than
the absolute abundance values among different systems. The efficiency
of the metal enrichment may depend on parameters such as age,
star-formation efficiency, and contribution from cD galaxies. However,
the relative contribution of SNe Ia and II and the process of metal
mixing in the ICM seem to be quite similar among different clusters and
groups.

\citet{tamura04} reported abundance ratios for 19 clusters studied
with XMM-Newton, and the mean Si/Fe ratio in cool and medium
temperature clusters with $kT < 6$~keV was $\sim 1.4$.  This is
consistent with the Suzaku results for groups and poor clusters
including A~262.  Their O/Fe ratio, $\sim0.6$, in the cluster core
also agrees with the Suzaku results including our A~262 case.
\citet{matsushita03} and \citet{matsushita07a} reported abundance
ratios for M87 and the Centaurus cluster, respectively, based on
XMM-Newton observations.  M87 showed Mg/O ratio to be $\sim 1.3$ in
the central region, and the Centaurus cluster indicated O/Fe and Si/Fe
ratios within $8'$ to be $\sim0.5$ and $\sim 1.0$, respectively, 
and they were consistent with our results.
\citet{sanders06} also showed the abundance ratios for the Centaurus 
cluster with Chandra and XMM-Newton, and the radial abundance ratios 
of O/Fe, Mg/Fe, and Si/Fe to be $0.5-1.0$, $\sim1.0$, $\sim1.2$, 
respectively, were consistent with our results.

\subsection{Number Ratio of SNe~II to SNe~Ia}

\begin{figure}
\centerline{\FigureFile(0.45\textwidth,8cm){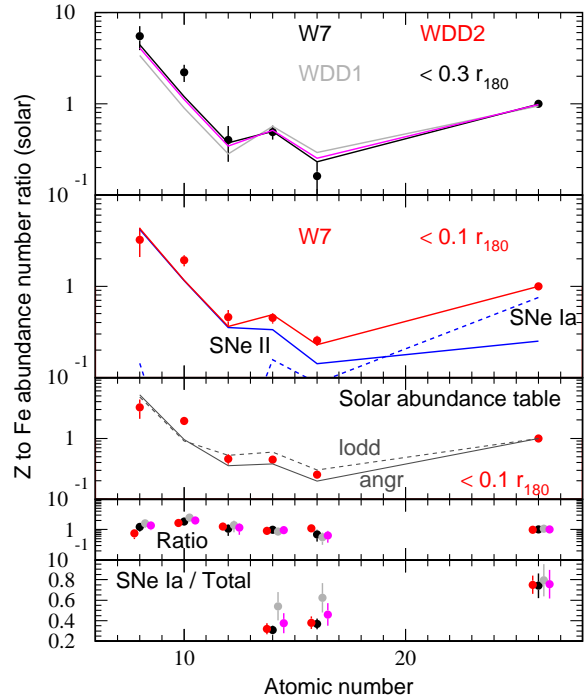}}
\vspace*{-1ex}
\caption{
Fit results of number ratios of elements to Fe of A~262.
Top and second panel show the abundance number ratios 
in solar unit within $0.27\;r_{\rm 180}$ (black) 
and $0.1\;r_{\rm 180}$ (red) regions, respectively.
Blue dashed and solid lines in the second panel correspond to 
the contributions of SNe~Ia (W7) and SNe~II within $0.1\;r_{180}$, 
respectively.
Ne (atomic number = 10) is excluded in the fit.
Third panel shows the comparison with solar abundance table of 
\citet{anders89} (light-gray line) and \citet{lodders03} (dashed
 light-gray line). 
Forth panel indicates ratios of data points to the best-fit model.
Bottom panel indicates fractions of the SNe~Ia contribution 
to total metals in the best-fit model for each element, respectively.
}\label{fig:6}
\end{figure}

\begin{table*}
\caption{
Integrated number of SNe~I ($N_{\rm Ia}$) and
number ratio of SNe~II to SNe~Ia ($N_{\rm II}$/$N_{\rm Ia}$).
}\label{tab:5}
\begin{center}
\begin{tabular}{lcccr}
\hline
\multicolumn{1}{c}{Region} &
\makebox[0in][c]{SNe~Ia Model} &
$N_{\rm Ia}$& $N_{\rm II}$/$N_{\rm Ia}$ & $\chi^2$/dof \\
\hline
 $<0.1\;r_{180}$ &W7 & $4.4\pm0.4\times10^{8}$& $3.0\pm 0.6$ & 8.6/3\\
 $< 0.27\;r_{\rm 180}$ &W7 & $2.7\pm0.9\times10^{9}$& $3.2\pm 1.0$ & 4.1/3\\
 $< 0.1\;r_{\rm 180}$ &WDD1 & $5.1\pm0.4\times10^{8}$& $1.8\pm 0.5$ & 27.2/3\\
 $< 0.27\;r_{\rm 180}$ &WDD1 & $2.9\pm1.1\times10^{9}$& $2.1\pm 0.9$ & 19.6/3\\
 $< 0.1\;r_{\rm 180}$ &WDD2 & $4.2\pm0.4\times10^{8}$& $2.9\pm 0.6$ & 9.2/3\\
 $< 0.27\;r_{\rm 180}$ &WDD2 & $2.5\pm0.9\times10^{9}$& $3.1\pm 1.0$ & 7.6/3\\
\hline
\end{tabular}
\end{center}
\end{table*}

In order to examine the SNe~Ia and SNe~II contribution to the ICM
metals, the elemental abundance pattern of O, Mg, Si, S and Fe was
examined for the region within 0.1 and $0.27~r_{180}$. The abundance
ratios at $r\sim0.3~r_{180}$ and $0.3<r<0.4~r_{180}$, with the latter
corresponding to the outermost region, look similar within the
statistic errors.  The relative abundance ratios to Fe were fitted
by a combination of average SNe~Ia and SNe~II yields per supernova, as
shown in figure~\ref{fig:6}.  The fit parameters were chosen to be the
integrated number of SNe~Ia ($N_{\rm Ia}$) and the number ratio of
SNe~II to SNe~Ia ($N_{\rm II}/N_{\rm Ia}$), because $N_{\rm Ia}$ could
be well constrained by the relatively small errors in the Fe
abundance.  The SNe Ia and II yields were taken from \citet{iwamoto99}
and \citet{nomoto06}, respectively.  We assumed a Salpeter IMF for
stellar masses from 10 to 50 $M_{\odot}$ with the progenitor
metallicity of $Z=0.02$ for SNe~II, and W7, WDD1 or WDD2 models for
SNe~Ia.  Table~\ref{tab:5} and figure~\ref{fig:6} summarizes the fit
results.  The number ratios were better represented by the W7 SNe~Ia
yield model than by WDD1\@. The number ratio of SNe~II to SNe~Ia with
W7 is $\sim3.2$ within 0.3~$r_{180}$, while the ratio with WDD1 is
$\sim 2.1$.  The WDD2 model gave the result quite similar to those of
W7\@.  The resultant number ratios are consistent with those in
\citet{sato07a}\@.  We also compared the abundance pattern of A~262
with the solar abundance.  The third panel in figure~\ref{fig:6} shows
the comparison of the abundance pattern within $r<0.1~r_{180}$ of
A~262 with the solar abundance pattern in \citet{anders89} and
\citet{lodders03}.  Our results of Mg, Si, and S fall between
the abundance patterns of \citet{anders89} and \citet{lodders03}\@.

Almost $\sim80$\% of Fe and $\sim30$\% of Si and S were synthesized
by SNe Ia in the W7 model, as demonstrated in the bottom panel of
figure~\ref{fig:6}. The observed tendency of $\sim2$ keV cluster is
similar to those of $\sim3$ keV clusters in \citet{sato07a}\@.  The
values in table~\ref{tab:5} imply that the $N_{\rm II}/N_{\rm Ia}$
ratio behaves in a similar manner for different supernova models
between $r<0.1~r_{180}$ and $r>0.1~r_{180}$.  We note that the fit was
not acceptable based on the $\chi^2$ value in table~\ref{tab:5}\@. As
described in \citet{sato07a}, the models adapted here (SNe yeild,
Salpter IMF, etc.) are probably too simplified.

\subsection{Metal Mass-to-Light Ratio}

\begin{table*}
\caption{
Integrated mass-to-light ratios of O, Mg, and Fe (OMLR, MMLR, IMLR ) 
with K-band luminosity in units of $M_{\odot}/L^{\rm K}_{\odot}$.
}\label{tab:6}
\begin{center}
\begin{tabular}{lccc}
\hline
\multicolumn{1}{c}{Region (kpc/$r_{180}$)} & OMLR & MMLR & IMLR \\
\hline
$<$59.8/0.05 & 4.5$^{+2.2}_{-2.1}\times10^{-4}$ &
 1.1$^{+0.2}_{-0.1}\times10^{-4}$ & 2.6$^{+0.1}_{-0.1}\times10^{-4}$ \\
$<$119.5/0.10 & 2.4$^{+0.7}_{-0.7}\times10^{-3}$ &
 3.4$^{+0.5}_{-0.5}\times10^{-4}$ & 7.4$^{+0.3}_{-0.3}\times10^{-4}$ \\
$<$179.3/0.14& 4.2$^{+0.9}_{-1.4}\times10^{-3}$ &
 7.4$^{+1.0}_{-1.0}\times10^{-4}$ & 1.4$^{+0.5}_{-0.5}\times10^{-3}$ \\
$<$338.6/0.27& 1.6$^{+1.0}_{-1.0}\times10^{-2}$ &
 1.2$^{+0.6}_{-0.5}\times10^{-3}$ & 3.0$^{+0.4}_{-0.3}\times10^{-3}$ \\
$<$537.8/0.43& 3.6$^{+2.4}_{-2.4}\times10^{-2}$ &
 2.4$^{+1.4}_{-1.3}\times10^{-3}$ & 4.6$^{+0.8}_{-0.7}\times10^{-3}$ \\
\hline
\end{tabular}
\end{center}
\end{table*}

\begin{figure*}
\centerline{
\FigureFile(0.45\textwidth,1cm){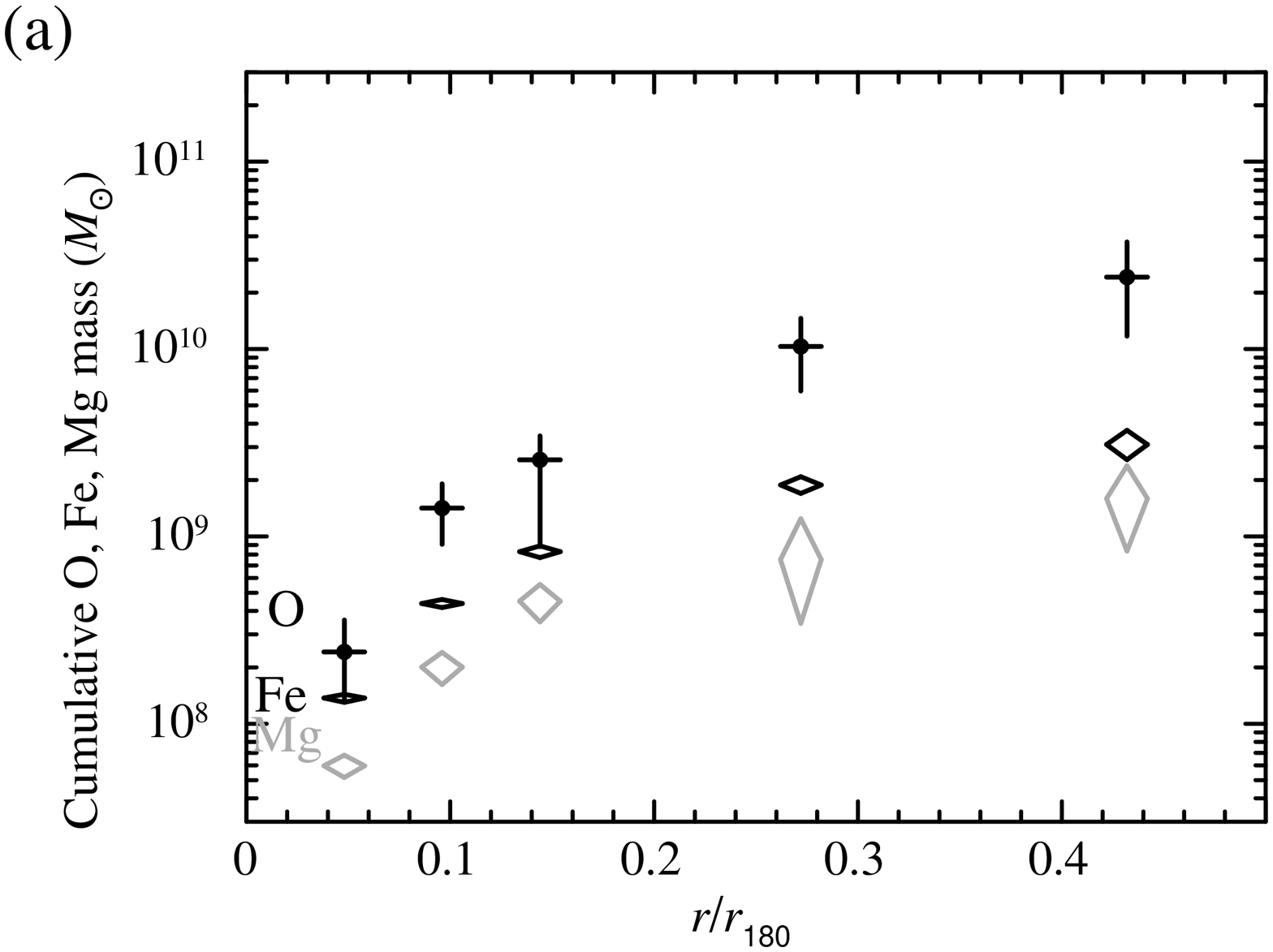}%
\hspace*{0.05\textwidth}
\FigureFile(0.45\textwidth,1cm){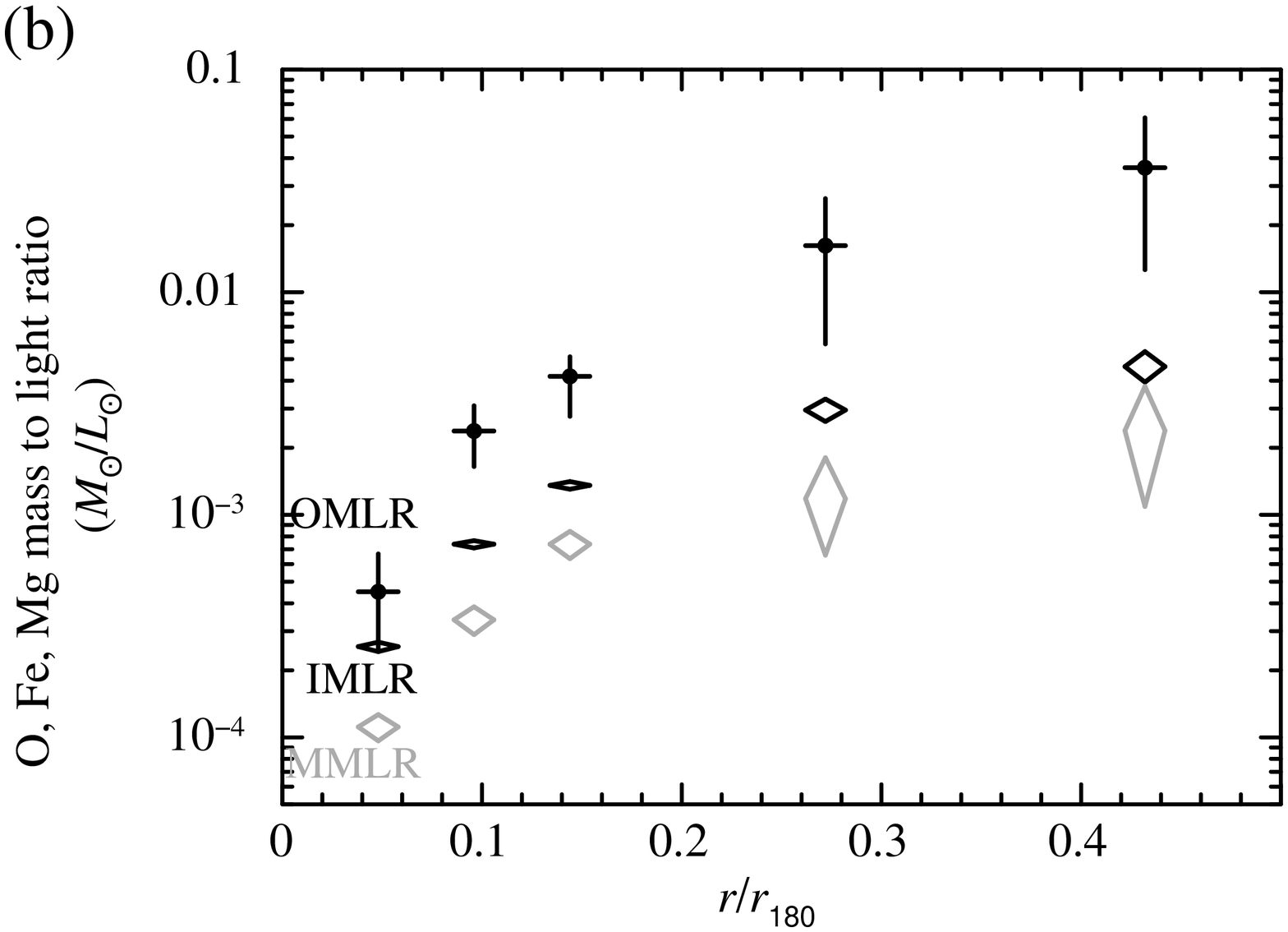}%
}
\caption{
(a) Cumulative mass, $M(<R)$, within the 3-dimensional radius, $R$,
for O, Fe, and Mg in A~262,
based on the combination of the abundance determination with Suzaku
and the gas mass profile with Chandra and XMM-Newton.
(b) Ratio of the O, Fe, and Mg mass in units of $M_\odot$
to the $K$ band luminosity in units of $L_\odot$
(OMLR, IMLR, and MMLR, respectively) against
the 3-dimensional radius. The black crosses, black diamonds, 
and gray diamonds show the OMLR, IMLR, MMLR, respectively. 
}\label{fig:7}
\end{figure*}

Combining the abundance profile obtained with Suzaku and the 3-dimensional
gas mass profile with Chandra and XMM-Newton \citep{gastaldello07}, 
we calculated cumulative metal mass as shown in figure~\ref{fig:7}(a).  
In order to see whether the gas mass profile 
derived from Chandra/XMM observations could apply to the offset regions 
of Suzaku which were out of the field of view of Chandra/XMM observations,
we compared three surface brightness profiles:
1.\ the Chandra/XMM observations which were used to derive the mass profiles,
2.\ the ASCA observations which were used to calculate ARFs, and
3.\ the present Suzaku observation.
However, we could not compare these three profiles directly 
because of the poor angular resolution and vignetting effect 
of Suzaku as shown in \citet{sato07a}. Thus, we need to simulate 
the surface brightness profiles using the parameters of 
the XMM/Chandra and ASCA observations with ``xissim'' Ftools task.
As a result, those profiles were consistent with that  of this Suzaku 
observation within $r<13'$, or ~260 kpc.
On the other hand, the profile in $r>20'$ looked slightly different, 
and the observed Suzaku profile was slightly steeper than those of 
Chandra/XMM and ASCA\@. Thus, the derived metal mass has a 
systematic error by about a factor of 2.
The derived iron, oxygen, and magnesium mass within
the 3-dimensional radius of $r< 540$~kpc are
$3.1\times 10^{9}$, $2.4\times 10^{10}$, and $1.6\times 10^{9}$ $M_\odot$,
respectively. 
Errors of the metal mass plotted in figure~\ref{fig:7}(a) are 
taken from the statistical errors of each elemental abundance 
in the spectral fits, because the statistical errors of the fits 
are lager than those of the gas mass profiles in \citet{gastaldello07}. 

We examined metal mass-to-light ratios for oxygen, iron, and magnesium
(OMLR, IMLR, and MMLR, respectively) to compare the ICM metal
distribution with the stellar mass profile.  Although, historically,
B-band luminosity has been used for the estimation of the stellar mass
\citep{makishima01}, we calculated the K-band luminosity in A~262
based on the Two Micron All Sky Survey \footnote{ The database
address: {\tt http://www.ipac.caltech.edu/2mass/}} with the Galactic
extinction, $A_K=0.032$, from NASA/IPAC Extragalactic Database (NED)
in the direction of A~262\@.  We used $2^{\circ}\times2^{\circ}$ data
set centered at the A~262 position as shown in table~\ref{tab:1}, and
subtracted the $r>1^{\circ}$ region as the background. In addition, we
deprojected the luminosity profile as a function of radius. 
The resultant luminosity within this Suzaku observation, $r<27'$, is
$6.7\times10^{11}~L_{\odot}$ in K-band. The radial luminosity profile
is not directly plotted, but it can be inferred from the ratio of the
metal mass in figure~\ref{fig:7}(a) divided by the metal mass to
light ratio in (b).  We calculated the integrated values of OMLR,
IMLR, and MMLR with K-band within $r\lesssim 540$~kpc as shown in
figure~\ref{fig:7}(b) and table~\ref{tab:6}, and their values turned
out to be $\sim 3.6\times 10^{-2}$, $\sim 4.6\times 10^{-3}$, and
$\sim 2.4\times 10^{-3}$ $M_{\odot}/L_{\odot}$, respectively.  The
errors are only based on the statistical errors of metal abundance in
the spectral fit, and the uncertainties of the gas mass profile and
the luminosity of member galaxies are not included.
Note that the derived luminosity profiles were not corrected 
the Suzaku PSF effect, because the errors of MLRs were dominated 
by the error of the metal mass compared with that of the light luminosity.

\begin{table*}
\caption{Comparison of IMLR, OMLR and MMLR with B-band luminosity 
for all systems.
}\label{tab:7}
\begin{center}
\begin{tabular}{lrrcrrl}
\hline\hline
& IMLR & OMLR & MMLR & \multicolumn{1}{c}{$r$ (kpc/$r_{180}$)} & \multicolumn{1}{c}{$k\langle T \rangle$} & Reference \\
\hline
Suzaku & & & & \\
NGC~720 $\dotfill$   & $1\times 10^{-4}$ & $4\times 10^{-4}$ & -- &25/0.04 & $\sim 0.56$ keV &       \citet{tawara08} \\
NGC~5044 $\dotfill$  & $2.6\times 10^{-3}$ & $6.6\times 10^{-3}$ &$1.6\times 10^{-3}$  &88/0.10 & $\sim 1.0$ keV & \\
                    & $3.6\times 10^{-3}$ & $9.4\times 10^{-3}$
 &$2.6\times 10^{-3}$  &260/0.30 &  &      \citet{komiyama08}\\
                    & $6.0\times 10^{-4}$ & -- & --  &327/0.38 &  &\citet{buote04}\\
Fornax $\dotfill$  & $4\times 10^{-4}$ & $2\times 10^{-3}$ & -- &130/0.13 & $\sim 1.3$ keV &         \citet{matsushita07b} \\
NGC~507 $\dotfill$  & $6.0\times 10^{-4}$ & $2.6\times 10^{-3}$ &$3.7\times 10^{-4}$  &120/0.11 & $\sim 1.5$ keV & \\
                    & $1.7\times 10^{-3}$ & $6.6\times 10^{-3}$ &$1.1\times 10^{-3}$  &260/0.24 &  &      \citet{sato08b}\\
HCG~62 $\dotfill$  & $2.0\times 10^{-3}$ & $6.4\times 10^{-3}$ &$1.0\times 10^{-3}$  &120/0.11 & $\sim 1.5$ keV & \\
                   & $4.6\times 10^{-3}$ & $3.8\times 10^{-2}$ &$1.5\times 10^{-3}$  &230/0.21 & &       \citet{tokoi08} \\
A~262 $\dotfill$   & $3.6\times 10^{-3}$ & $1.2\times 10^{-2}$ &$1.6\times 10^{-3}$  &130/0.10 & $\sim 2$ keV & \\
                   & $6.7\times 10^{-3}$ & $3.7\times 10^{-2}$ &$2.7\times 10^{-3}$  &340/0.27 & &      This work \\
A~1060 $\dotfill$   & $5.7\times 10^{-3}$ & $4.3\times 10^{-2}$ &$2.4\times 10^{-3}$  &180/0.12 & $\sim 3$ keV &   \\
                    & $4.0\times 10^{-3}$ & $4.3\times 10^{-2}$ &$1.6\times 10^{-3}$  &380/0.25 & &        \citet{sato07a} \\
AWM~7 $\dotfill$   & $4.8\times 10^{-3}$ & $2.6\times 10^{-2}$ &$3.4\times 10^{-3}$  & 180/0.11 & $\sim 3.5$ keV& \\
                   & $7.6\times 10^{-3}$ & $3.1\times 10^{-2}$ &$6.7\times 10^{-3}$  & 360/0.22 & &       \citet{sato08a}\\
\hline
XMM-Newton & & & & \\
Centaurus $\dotfill$ & $4\times 10^{-3}$ & $3\times 10^{-2}$ &-- &190/0.11 & $\sim 4$ keV &        \citet{matsushita07a} \\
\hline
\end{tabular}
\end{center}
\end{table*}

\begin{figure*}
\centerline{
\FigureFile(0.45\textwidth,1cm){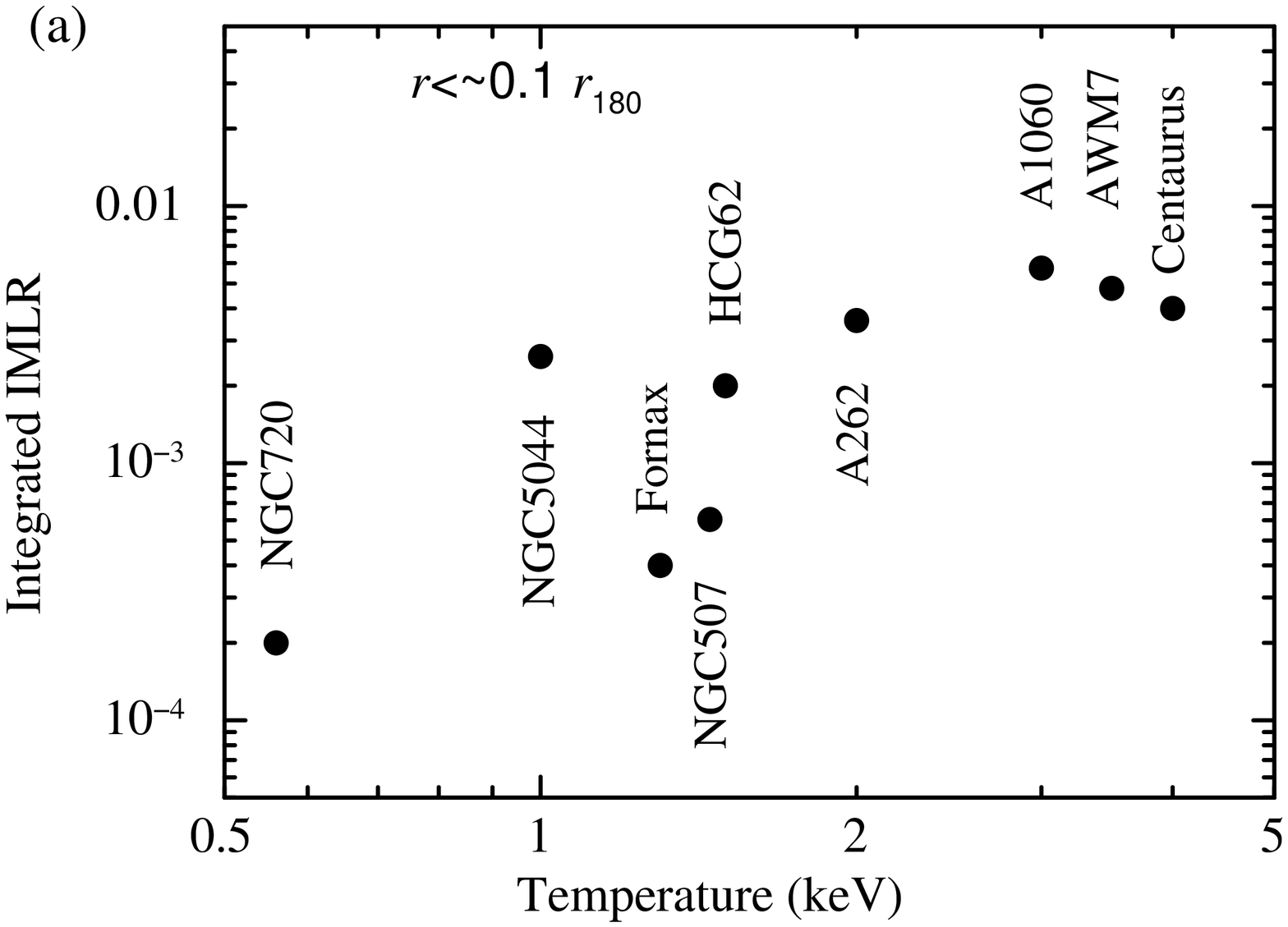}%
\hspace*{0.05\textwidth}
\FigureFile(0.45\textwidth,1cm){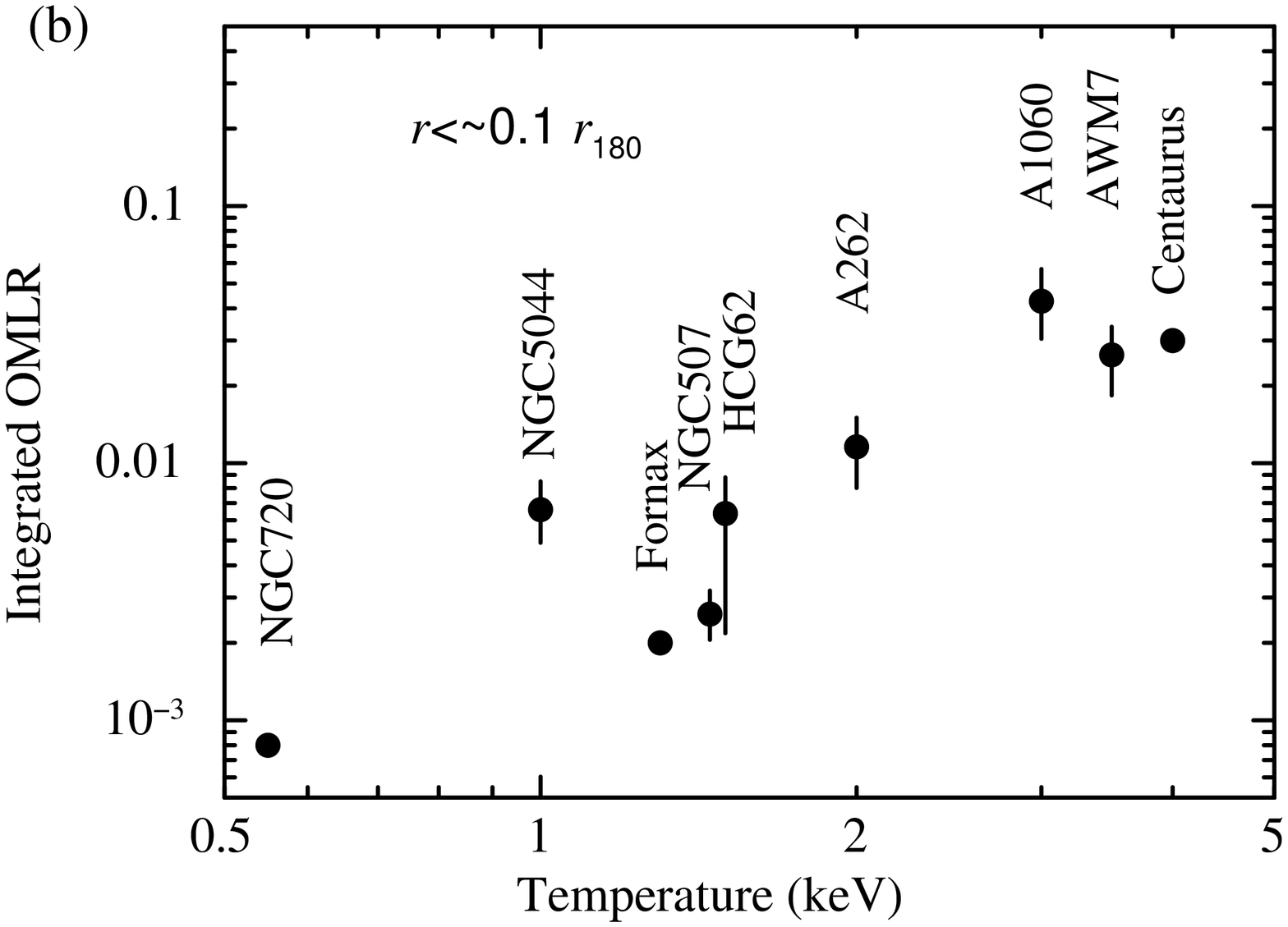}%
}
\centerline{
\FigureFile(0.45\textwidth,1cm){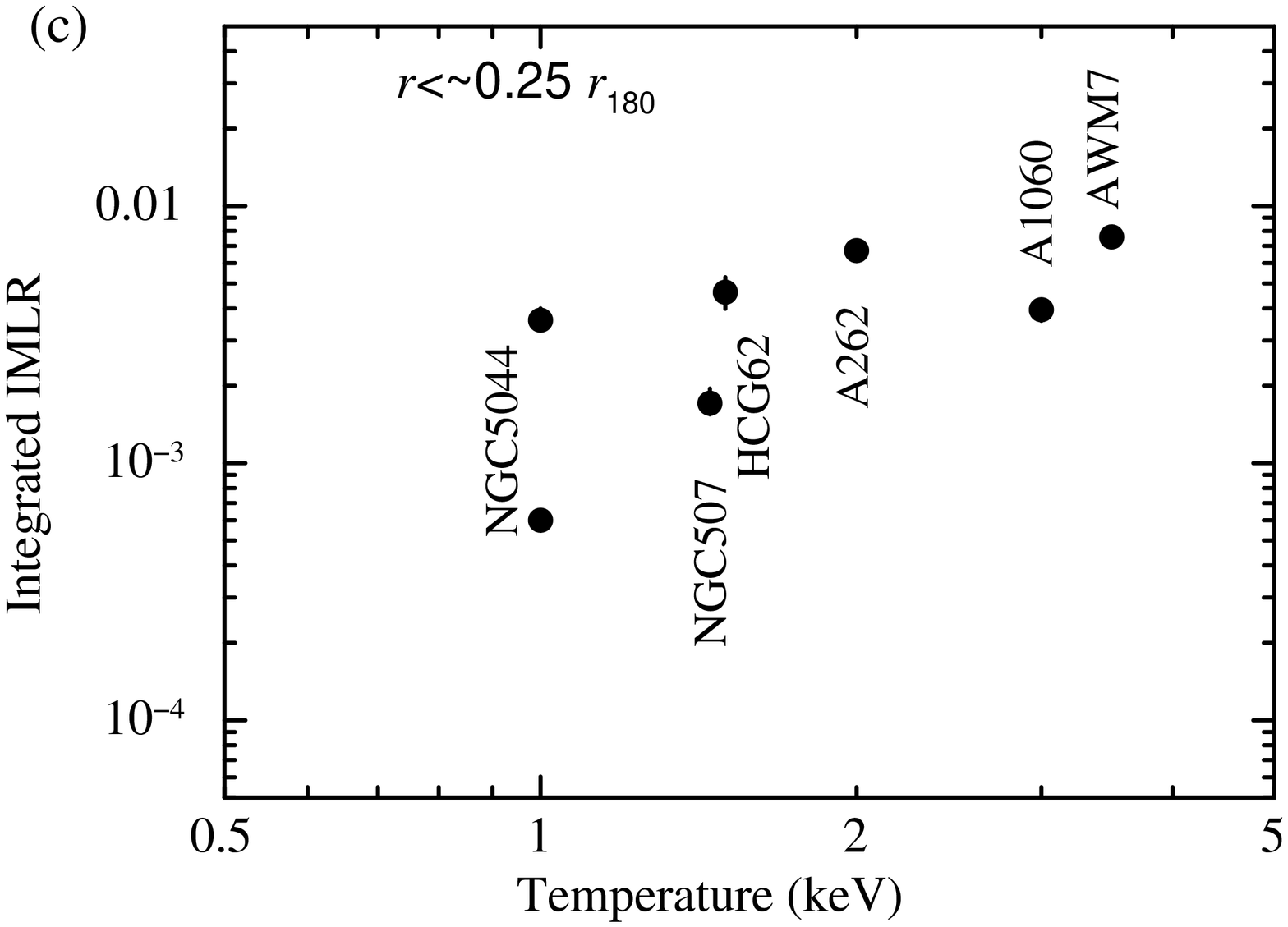}%
\hspace*{0.05\textwidth}
\FigureFile(0.45\textwidth,1cm){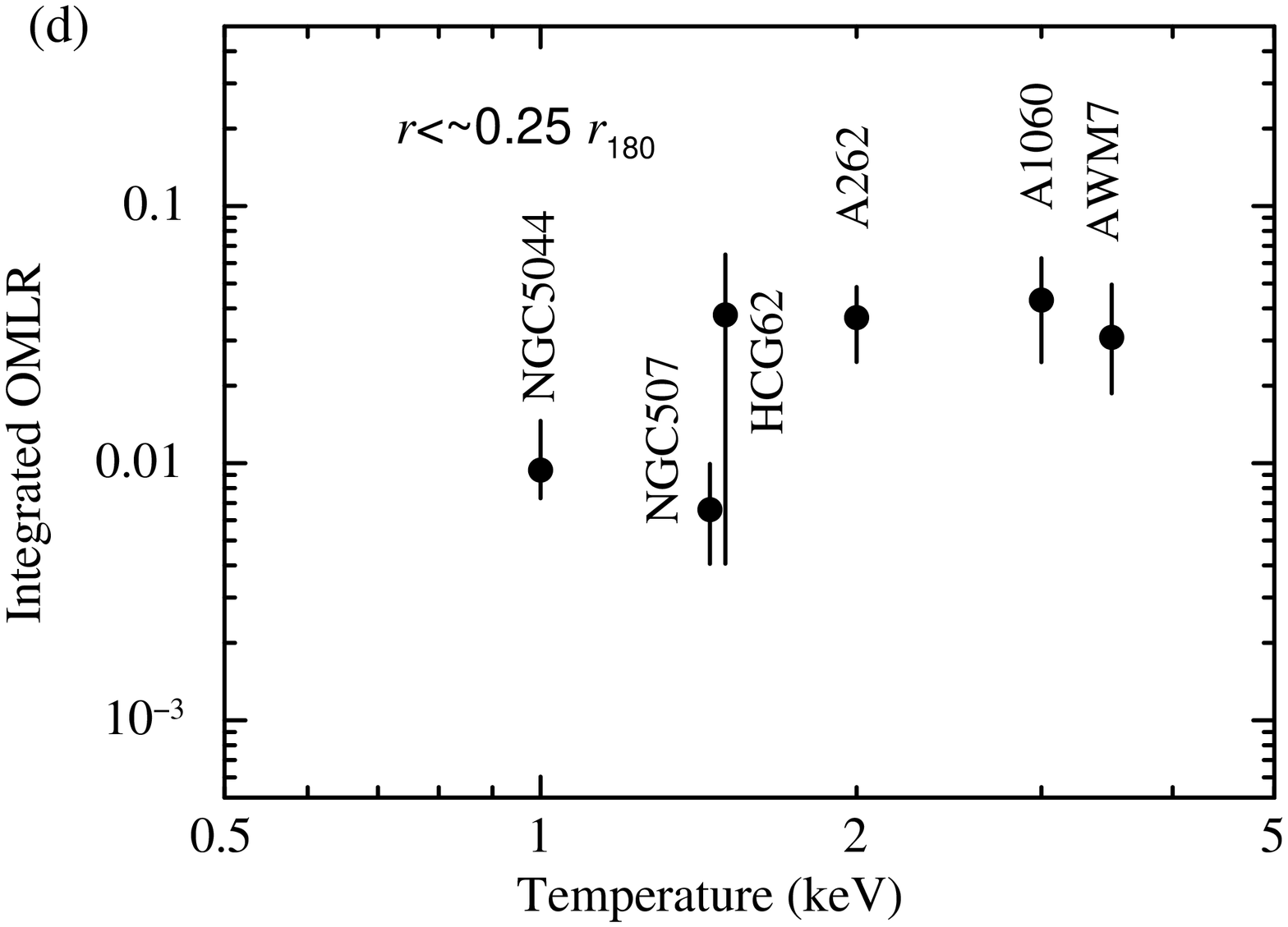}%
}
\caption{
Comparison of IMLR (a) and (c), and OMLR (b) and (d) 
with B-band luminosity to the other clusters and groups within 
$\sim0.1~r_{180}$ and $\sim0.25~r_{180}$ region, respectively. 
}\label{fig:8}
\end{figure*}

We also derived the MLRs with B-band luminosity using an appropriate
color $B-K=4.2$ for early-type galaxies given by \citet{lin04}, along
with the Galactic extinction, $A_B=0.267$, from NED in the direction
of A~262\@. The integrated values of OMLR, IMLR, and MMLR with B-band
within $r\lesssim 540$~kpc turned out to be $\sim 7.1\times 10^{-2}$,
$\sim 9.0\times 10^{-3}$, and $\sim 4.7\times 10^{-3}$
$M_{\odot}/L_{\odot}$, respectively. Comparing these B-band MLRs with
those of other clusters and groups measured within $\sim0.1~r_{180}$
and $\sim0.25~r_{180}$, the A~262 points fall between those of
clusters and groups, in particular within $\sim0.1~r_{180}$. In other
radii, the A262 results are consistent with those for the poor
clusters as shown in table~\ref{tab:7} and figure~\ref{fig:8}.  The
tendency that smaller systems with lower gas temperature tend to
exhibit smaller MLRs, as shown previously by \citet{makishima01} and
by \citet{sato08b}, is clearly indicated by the present addition of
the A~262 data.

\section{Summary and conclusion}

Based on the Suzaku observation of A~262, we studied spatial
distribution of temperature and metal abundances for O, Mg, Si, S, and
Fe up to $\sim0.43~r_{180}$\@. The ICM temperature decreases mildly
from $\sim2.3$ keV to $\sim1.5$ keV in the outer region.  The
abundances of Mg, Si, S, and Fe drop from subsolar levels at the center
to $\sim 1/5$ solar in the outermost region, while the O abundance
shows a flatter distribution around $\sim 0.5$ solar compared with the
other metals.  The abundance ratios, O/Fe, Mg/Fe, Si/Fe, and S/Fe for
A~262 are similar to those of other clusters with $kT = $ 3--4 keV and
groups with $kT \sim 1.5$ keV\@.  The number ratio of SNe II to Ia
which contributed for the ICM enrichment is $3.0 \pm 0.6$, consistent
with the values for other clusters and groups. The derived MLRs with
B-band luminosity fall between the data for clusters and groups
measured within a radius of $\sim0.1~r_{180}$.  Thus, the A~262 data
gives us a good connection concerning the global picture of the metal
enrichment process between clusters and groups of galaxies.

\bigskip
Authors are grateful to T.~Ohashi for valuable comments and discussions.
We also thank the referee for providing valuable comments.
Part of this work was financially supported by the Ministry of
Education, Culture, Sports, Science and Technology of Japan,
Grant-in-Aid for Scientific Research
Nos.\ 18740011, 19840043.

This research has made use of the NASA/IPAC Extragalactic Database 
(NED) which is operated by the JPL, under contract with NASA.


\begin{thebibliography}{}
\bibitem[Anders \& Grevesse(1989)]{anders89}
        Anders, E., \& Grevesse, N.\ 1989, \gca, 53, 197 

\bibitem[Arimoto et al.(1997)]{arimoto97} Arimoto, N., 
Matsushita, K., Ishimaru, Y., Ohashi, T., 
\& Renzini, A.\ 1997, \apj, 477, 128 

\bibitem[Arnaud et al.(1992)]{arnaud92} Arnaud, M., Rothenflug, 
R., Boulade, O., Vigroux, L., \& Vangioni-Flam, E.\ 1992, \aap, 254, 49 

\bibitem[Blanton et al.(2004)]{blanton04} Blanton, E.~L., 
Sarazin, C.~L., McNamara, B.~R., \& Clarke, T.~E.\ 2004, \apj, 612, 817 

\bibitem[B{\"o}hringer et al.(2005)]{boehringer05} B{\"o}hringer, 
H., Matsushita, K., Finoguenov, A., Xue, Y., \& Churazov, E.\ 2005, 
Advances in Space Research, 36, 677 

\bibitem[Buote(2000)]{buote00} Buote, D.~A.\ 2000, \mnras, 311, 
176 

\bibitem[Buote et al.(2004)]{buote04} Buote, D.~A., Brighenti, 
F., \& Mathews, W.~G.\ 2004, \apjl, 607, L91 

\bibitem[David et al.(1993)]{david93} David, L.~P., Slyz, A., 
Jones, C., Forman, W., Vrtilek, S.~D., 
\& Arnaud, K.~A.\ 1993, \apj, 412, 479 

\bibitem[David et al.(1996)]{david96} David, L.~P., Jones, C., 
\& Forman, W.\ 1996, \apj, 473, 692 

\bibitem[De Grandi \& Molendi(2001)]{degrandi01} 
De Grandi, S., \& Molendi, S.\ 2001, \apj, 551, 153 

\bibitem[de Plaa et al.(2007)]{deplaa07} de Plaa, J., Werner,
N., Bleeker, J.~A.~M., Vink, J., Kaastra, J.~S., \& M{\'e}ndez, M.\ 2007,
\aap, 465, 345

\bibitem[de Plaa et al.(2006)]{deplaa06} de Plaa, J., et al.\
2006, \aap, 452, 397

\bibitem[Dickey \& Lockman(1990)]{dickey90}
        Dickey, J.~M., \& Lockman, F.~J.\ 1990, \araa, 28, 215 

\bibitem[Fabian et al.(2006)]{fabian06} Fabian, A.~C., Sanders, 
J.~S., Taylor, G.~B., Allen, S.~W., Crawford, C.~S., Johnstone, R.~M., 
\& Iwasawa, K.\ 2006, \mnras, 366, 417 


\bibitem[Fabian et al.(2005)]{fabian05} Fabian, A.~C., Sanders, 
J.~S., Taylor, G.~B., \& Allen, S.~W.\ 2005, \mnras, 360, L20 


\bibitem[Fabian et al.(2001)]{fabian01} Fabian, A.~C., Sanders, 
J.~S., Ettori, S., Taylor, G.~B., Allen, S.~W., Crawford, C.~S., Iwasawa, 
K., \& Johnstone, R.~M.\ 2001, \mnras, 321, L33 

\bibitem[Fanti et al.(1986)]{fanti86} Fanti, C., Fanti, R., de 
Ruiter, H.~R., \& Parma, P.\ 1986, \aaps, 65, 145 

\bibitem[Finoguenov et al.(2000)]{finoguenov00}
        Finoguenov, A., David, L.~P., \& Ponman, T.~J.\ 2000, \apj, 544, 188 

\bibitem[Finoguenov et al.(2001)]{finoguenov01}
        Finoguenov, A., Arnaud, M., \& David, L.~P.\ 2001, \apj, 555, 191 

\bibitem[Finoguenov et al.(2002)]{finoguenov02} Finoguenov, A., 
Matsushita, K., B{\"o}hringer, H., Ikebe, Y., \& Arnaud, M.\ 2002, \aap, 
381, 21 

\bibitem[Fujimoto et al.(2007)]{fujimoto07} Fujimoto, R., et al.\ 
2007, \pasj, 59, 133 

\bibitem[Fukazawa et al.(1998)]{fukazawa98}
        Fukazawa, Y., Makishima, K., Tamura, T., Ezawa, H., Xu, H.,
        Ikebe, Y., Kikuchi, K., \& Ohashi, T.\ 1998, \pasj, 50, 187 

\bibitem[Fukazawa et al.(2000)]{fukazawa00}
        Fukazawa, Y., Makishima, K., Tamura, T., Nakazawa, K.,
        Ezawa, H., Ikebe, Y., Kikuchi, K., \& Ohashi, T.\ 2000, \mnras, 313, 21

\bibitem[Fukazawa et al.(2004)]{fukazawa04} Fukazawa, Y., 
Makishima, K., \& Ohashi, T.\ 2004, \pasj, 56, 965 

\bibitem[Gastaldello \& Molendi(2002)]{gastaldello02} 
Gastaldello, F., \& Molendi, S.\ 2002, \apj, 572, 160 

\bibitem[Gastaldello et al.(2007)]{gastaldello07} Gastaldello, F., 
Buote, D.~A., Humphrey, P.~J., Zappacosta, L., Bullock, J.~S., Brighenti, 
F., \& Mathews, W.~G.\ 2007, \apj, 669, 158 

\bibitem[Gastaldello et al.(2008)]{gastaldello08} 
Gastaldello, F., et al.\ in preparation

\bibitem[Humphrey \& Buote(2006)]{humphrey06} Humphrey, P.~J., \& Buote, D.~A.\
2006, \apj, 639, 136 

\bibitem[Ishisaki et al.(2007)]{ishisaki07} Ishisaki, Y., et al.\ 
        2007, \pasj, 59, 113 

\bibitem[Iwamoto et al.(1999)]{iwamoto99} Iwamoto, K., Brachwitz,
F., Nomoto, K., Kishimoto, N., Umeda, H., Hix, W.~R., \& Thielemann, F.-K.\
1999, \apjs, 125, 439

\bibitem[Komiyama et al.(2008)]{komiyama08} Komiyama, M., et al.\ 
        2008, in preparation

\bibitem[Koyama et al.(2007)]{koyama07} Koyama, K., et al.\ 
        2007, \pasj, 59, 23 

\bibitem[Kushino et al.(2002)]{kushino02} Kushino, A., Ishisaki, 
Y., Morita, U., Yamasaki, N.~Y., Ishida, M., Ohashi, T., \& Ueda, Y.\ 2002, 
\pasj, 54, 327 

\bibitem[Lin \& Mohr(2004)]{lin04} Lin, Y.-T., \& Mohr, 
J.~J.\ 2004, \apj, 617, 879 

\bibitem[Lodders(2003)]{lodders03} Lodders, K.\ 2003, \apj, 591, 
1220 

\bibitem[Lumb et al.(2002)]{lumb02} Lumb, D.~H., Warwick, 
R.~S., Page, M., \& De Luca, A.\ 2002, \aap, 389, 93 

\bibitem[Makishima et al.(2001)]{makishima01}
        Makishima, K., et al.\ 2001, \pasj, 53, 401 

\bibitem[Markevitch et al.(1998)]{markevitch98}
        Markevitch, M., et al.\ 1998, \apj, 503, 77

\bibitem[Matsushita et al.(2000)]{matsushita00} Matsushita, K., 
Ohashi, T., \& Makishima, K.\ 2000, \pasj, 52, 685 

\bibitem[Matsushita et al.(2003)]{matsushita03} Matsushita, K., 
Finoguenov, A., B{\"o}hringer, H.\ 2003, \aap, 401, 443 

\bibitem[Matsushita et al.(2007a)]{matsushita07a} Matsushita, K., 
        B{\"o}hringer, H., Takahashi, I., \& Ikebe, Y.\ 2007b, \aap, 462, 953 

\bibitem[Matsushita et al.(2007b)]{matsushita07b} Matsushita, K., et 
al.\ 2007a, \pasj, 59, 327 

\bibitem[McCammon et al.(2002)]{mccammon02} McCammon, D., et al.\ 
2002, \apj, 576, 188 

\bibitem[Neill et al.(2001)]{neill01} Neill, J.~D., Brodie, 
J.~P., Craig, W.~W., Hailey, C.~J., \& Misch, A.~A.\ 2001, \apj, 548, 550 

\bibitem[Nomoto et al.(2006)]{nomoto06} Nomoto, K., Tominaga, 
N., Umeda, H., Kobayashi, C., \& Maeda, K.\ 2006, Nuclear Physics A, 777, 
424 

\bibitem[O'Sullivan et al.(2005)]{osullivan05} O'Sullivan, E., 
Vrtilek, J.~M., Kempner, J.~C., David, L.~P., 
\& Houck, J.~C.\ 2005, \mnras, 357, 1134 

\bibitem[Parma et al.(1986)]{parma86} Parma, P., 
de Ruiter, H.~R., Fanti, C.,  \& Fanti, R.\ 1986, \aaps, 64, 135 

\bibitem[Peres et al.(1998)]{peres98} Peres, C.~B., Fabian, 
A.~C., Edge, A.~C., Allen, S.~W., Johnstone, R.~M., 
\& White, D.~A.\ 1998, \mnras, 298, 416 

\bibitem[Peterson et al.(2003)]{peterson03} Peterson, J.~R., Kahn, 
S.~M., Paerels, F.~B.~S., Kaastra, J.~S., Tamura, T., Bleeker, J.~A.~M., 
Ferrigno, C., \& Jernigan, J.~G.\ 2003, \apj, 590, 207 

\bibitem[Rasmussen \& Ponman(2007)]{rasmussen07} Rasmussen, J., 
\& Ponman, T.~J.\ 007, \mnras, 380, 1554 

\bibitem[Renzini et al.(1993)]{renzini93} Renzini, A., Ciotti, 
L., D'Ercole, A., \& Pellegrini, S.\ 1993, \apj, 419, 52 

\bibitem[Renzini(1997)]{renzini97} Renzini, A.\ 1997, \apj, 488, 
35

\bibitem[Sanders \& Fabian(2002)]{sanders02} 
Sanders, J.~S., \& Fabian, A.~C.\ 2002, \mnras, 331, 273 

\bibitem[Sanders \& Fabian(2006)]{sanders06} 
Sanders, J.~S., \& Fabian, A.~C.\ 2006, \mnras, 371, 1483 

\bibitem[Sanders \& Fabian(2007)]{sanders07} 
Sanders, J.~S., \& Fabian, A.~C.\ 2007, \mnras, 381, 1381 

\bibitem[Simionescu et al.(2008)]{simionescu08} 
Simionescu, A., Werner, N., Finoguenov,
 A., B{\"o}hringer, H., \& Br{\"u}ggen,
 M.\ 2008, \aap, 482, 97 

\bibitem[Sato et al.(2007a)]{sato07a} Sato, K., et al.\ 
        2007a, \pasj, 59, 299 

\bibitem[Sato et al.(2007b)]{sato07b} Sato, K., Tokoi, K., 
Matsushita, K., Ishisaki, Y., Yamasaki, N.~Y., Ishida, M., \& Ohashi, T.\ 
2007, \apjl, 667, L41 

\bibitem[Sato et al.(2008a)]{sato08a} Sato, K., Matsushita, K., 
Ishisaki, Y., Yamasaki, N.~Y., Ishida, M., Sasaki, S., \& Ohashi, T.\ 2008,
\pasj, 60, S333 

\bibitem[Sato et al.(2008b)]{sato08b} Sato, K., Matsushita, K., 
Ishisaki, Y., Yamasaki, N.~Y., Ishida, M., \& Ohashi, T.\ 2008,
\pasj, in press

\bibitem[Tamura et al.(2003)]{tamura03} Tamura, T., Kaastra, 
J.~S., Makishima, K., \& Takahashi, I.\ 2003, \aap, 399, 497 

\bibitem[Tamura et al.(2004)]{tamura04} Tamura, T., Kaastra, 
J.~S., den Herder, J.~W.~A., Bleeker, J.~A.~M., \& Peterson, J.~R.\ 2004, 
\aap, 420, 135 

\bibitem[Tawara et al.(2008)]{tawara08} Tawara, Y., et al.\ 
        2008, \pasj, 60, S307

\bibitem[Tokoi et al.(2008)]{tokoi08} Tokoi, K., et al.\ 2008, 
\pasj, 60, S317

\bibitem[Vikhlinin et al.(2006)]{vikhlinin06} Vikhlinin, A., 
Kravtsov, A., Forman, W., Jones, C., Markevitch, M., Murray, S.~S., 
\& Van Speybroeck, L.\ 2006, \apj, 640, 691 

\bibitem[Werner et al.(2006)]{werner06} Werner, N., de Plaa, J., 
Kaastra, J.~S., Vink, J., Bleeker, J.~A.~M., Tamura, T., Peterson, J.~R., 
\& Verbunt, F.\ 2006a, \aap, 449, 475 

\bibitem[White(2000)]{white00} White, D.~A.\ 2000, \mnras, 312, 
663 

\bibitem[Xu et al.(2002)]{xu02} Xu, H., et al.\ 2002, \apj, 
579, 600 

\end{thebibliography}
\end{document}